# Ultraviolet Radiation from Evolved Stellar Populations
# I. Models

Ben Dorman, Robert T. Rood & Robert W. O'Connell

Dept. of Astronomy

University of Virginia, P.O.Box 3818, University Station, Charlottesville, VA 22903-0818, U.S.A.

*Abstract*

This series of papers comprises a systematic exploration of the hypothesis that the far ultraviolet radiation from star clusters and elliptical galaxies originates from extremely hot horizontal-branch (HB) stars and their post-HB progeny. This first paper presents an extensive grid of calculations of stellar models from the Zero Age Horizontal Branch through to a point late in post-HB evolution or a point on the white dwarf cooling track. The grid will be used to produce synthesized UV fluxes for the interpretation of existing and future short wavelength (900-3000 Å) observations. Our sequences have been computed for a range of masses which concentrates on models that begin their HB evolution very close to the hot end of the ZAHB. We have calculated tracks for three metal-poor compositions ([Fe/H] = $-2.26, -1.48, -0.47$ with [O/Fe] $> 0$), for use with globular cluster observations. We have also chosen three metal rich compositions ($Z = 0.017 = Z_\odot$, $Z = 0.04, 0.06$) for use in the study of elliptical galaxy populations. For each of the two super-metal-rich compositions, for which the helium abundance is unconstrained by observation, we have computed two sets of sequences: one assuming no additional helium, and a second with a large enhancement ($Y_{HB} = 0.29$ and 0.36 for $Z = 0.04$), and ($Y_{HB} = 0.29$ and 0.46 for $Z = 0.06$). For each set of sequences our lowest ZAHB envelope masses ($M^0_{env}$) are in the range $0.002 < M^0_{env} < 0.006 \, M_\odot$.

We use the term 'Extreme Horizontal Branch' (EHB) to refer to HB sequences of constant mass that do not reach the thermally-pulsing stage on the AGB. These models evolve after core helium exhaustion into Post-Early Asymptotic Giant Branch (AGB) stars, which leave the AGB before thermal pulsing, and AGB-Manqué stars, which never reach the AGB. We describe various features of the evolution of post-HB stars, discussing the correspondence between slow phases of evolution at high temperature and the Early-AGB evolution. We note that the relationship between core mass and luminosity for stars on the upper AGB is not straightforward, because stars arrive on the ZAHB with a range of masses and subsequently burn different amounts of fuel. We determine from our models an upper bound to the masses of EHB stars, finding that it varies little for [Fe/H] $< 0$, but that it is sensitive to the helium abundance. We show that for each composition there is a range of $M^0_{env}$ (at least a few hundredths $M_\odot$) in which the models have a slow phase of evolution at high temperature. The duration of this phase is found to increase with the metallicity, but its luminosity is lower, so that total UV energy output is not significantly different from metal-poor sequences. The properties of very metal rich stars are, however, made uncertain by our lack of knowledge of the helium abundance for [Fe/H] $> 0$; the range of stellar masses in which high temperatures are attained for significant periods of time increases with $Y$. There is no intrinsic composition dependence of the peak UV output from evolved stars; the output from a stellar population depends most directly on the mass distribution of stars arriving on the ZAHB. This is determined mainly by the mass loss that occurs on the red giant branch.

*Subject Headings:* galaxies: stellar content — stars: AGB and post-AGB — stars:evolution — stars:Population II — ultraviolet: galaxies

## *1  Introduction*

The evolution of hot horizontal-branch (HB) stars has become the focus of much interest over the last few years as space-based observations have begun to allow detailed studies of hot components of stellar populations. The OAO-2 (Welch and Code 1979), and ANS satellites (van Albada, deBoer and Dickens 1981) found evidence for a strong vacuum ultraviolet component in the radiation from some globular clusters. M 79 and M 62, which have horizontal branches that extend at least to the limit of detectability — down to and below the magnitude of the main-sequence turnoff — in optical wavebands, were also found to have significant UV flux at 1500 Å(see also Caloi, Castellani & Piccolo 1987). A larger set of globular clusters also contain a small number of very bright UV sources (Zinn, Newell, & Gibson 1972; deBoer 1985; Landsman *et al.* 1992), which are identified with the penultimate stages of low-mass stellar evolution. The sdB and sdOB stars (Greenstein & Sargent 1974; Green, Schmidt & Leibert 1986) are also thought to be extreme HB objects, with hydrogen-rich envelopes of 0.003 – 0.02 $M_\odot$.

The OAO-2 satellite (Code 1969) also discovered a high temperature stellar component in the UV radiation from elliptical galaxies and spiral galaxy bulges. This "UV upturn phenomenon", or "UVX," is regarded as an important clue to the stellar content and spectral evolution of elliptical galaxies and spiral bulges. The upturn in the spectra is found in nearly all ellipticals, and is produced by hotter stars than those in the globular clusters. Evolved helium-burning stars of various types have become prime candidates for the origin of this flux, as discussed by Greggio & Renzini (1990, hereafter GR90) and O'Connell (1993). Considerable theoretical activity has been generated by the intriguing observation (Faber 1983; Burstein *et al.* 1988; Ferguson & Davidsen 1993) that the UVX may be positively correlated with the galaxies' metal abundance. Understanding the origin and evolution of this hot component may provide an important diagnostic for the evolution of the galaxies themselves, as clues to the age, metal abundance and other properties of the old populations. The hot phases are especially useful since they can be readily observed in the integrated Far-UV light of galaxies without serious confusion from the cooler stars that dominate the optical and infra-red spectrum.

The properties of stellar models in post-HB and post-Asymptotic Giant Branch evolutionary phases have been investigated by Paczyński (1971), Gingold (1976), Schönberner (1979, 1983), Wood and Faulkner (1986), Caloi (1989), Castellani & Tornambè (1991), Castellani, Limongi & Tornambè (1992), and Horch, Demarque & Pinnsoneault (1992, hereafter HDP). Prior to the helium-burning stages of evolution, stars with initial zero age main sequence mass $M_{\mathrm{ZAMS}} \lesssim 2.2 M_\odot$ evolve to the tip of the red-giant branch (RGB) where they undergo a violent ignition of helium burning known as the helium core flash. At this stage, the hydrogen burning shell has produced a central helium core of about 0.5 $M_\odot$ in which $Y = 1 - Z$. During the flash and relaxation to quiescent core helium burning on the Zero-Age Horizontal Branch (ZAHB) the carbon abundance of the core increases by around 3% but the core mass remains essentially unchanged. The core mass characterizing a ZAHB is $M_c^0$, assumed to vary little in a population of stars of fixed composition. Studies of HB morphology in globular clusters and observations of the field HB population imply that RGB stars lose a substantial amount of mass $\delta M_{RGB} \sim 0.1 - 0.3 M_\odot$ prior to arriving on the HB with total mass $M_{\mathrm{HB}} = M_{\mathrm{ZAMS}} - \delta M_{\mathrm{RGB}}$ (*cf.* Iben & Rood 1970; Renzini 1981b).

The size of the envelope mass at the ZAHB ($M_{\mathrm{env}}^0 = M_{\mathrm{HB}} - M_c^0$) is the key parameter that determines the properties of the He-burning stellar models. For $M_{\mathrm{HB}} \lesssim 1 M_\odot$, the surface temperature increases as $M_{\mathrm{env}}^0$ is decreased, the point $M_{\mathrm{env}}^0 = 0$ being marked by the helium main-sequence star with $M = M_c^0$. We use the symbol $M_{\mathrm{env}}$ to refer to the envelope mass during evolution. During the HB evolution the luminosity of the models depends largely on $M_{\mathrm{env}}^0$, which determines the strength of the hydrogen-burning shell. The core mass $M_c$ thus increases to a greater or lesser degree depending on $M_{\mathrm{env}}^0$; for the least massive models, $M_c$ changes hardly at all before core helium exhaustion. In the ensuing evolution from the HB to the tip of the AGB the bulk of the energy is supplied by hydrogen burning, and $M_c$ grows much more rapidly as a result. There is a critical $M_{\mathrm{env}}^0 = M_{\mathrm{env}}^{\mathrm{TP}} \sim 0.05 M_\odot$, above which the fuel supply is adequate to enable the evolution to reach the later stages of AGB evolution. According to our current understanding, the evolution on the thermally pulsing AGB (TP-AGB) culminates, after a number of pulses (also called "shell flashes") with a period of rapid mass loss or superwind (Renzini 1981a), in the formation of a planetary nebula. During this period, enough of the envelope is removed that the star can no longer sustain a convective exterior, and it evolves rapidly to higher temperatures while retaining fixed luminosity. For models with $M_{HB} > M_{env}^{TP} + M_c^0$ the hydrogen fuel consumed during the HB phase increases with $M_{env}^0$, while that burnt in reaching the upper AGB varies less with the mass. The mass of the core at a given luminosity in the post-AGB stages tends to be an increasing function of $M_{HB}$. The length of time spent as a hot, UV-producing object depends upon the luminosity, upon the degree to which mass loss still operates, and upon whether the helium burning shell is active as the star leaves the AGB. The latter depends, in turn, on whether the blueward evolution is triggered during a He shell flash, in which the hydrogen burning luminosity is negligible, or starts in a quiescent interpulse phase, in which the hydrogen shell is dominant (see Wood & Faulkner 1986; Iben *et al.* 1983). In any case, however, the time spent at high temperature in this Post-AGB phase is shorter than for their lower mass, more exotic counterparts described below.

Figure 1 illustrates schematically the various possibilities for post-HB evolution. The normal AGB sequence described above is shown as a solid line. Its evolution through a thermal pulsing stage to a high post-AGB luminosity $\log L/L_\odot \gtrsim 3.5$ is indicated. The region of the HR diagram in which the rates of mass loss are greatest is also marked.

If the ZAHB envelope mass is less than some critical mass plus an amount to allow for mass loss ($M_{env}^0 < M_{env}^{TP} + \delta M_{AGB}$), the model will never become a "classic" TP-AGB star. We will refer to these as Extreme HB objects (EHB). Note that the EHB as we define it is determined by the post-HB evolutionary behavior of the models. Our EHB classification encompasses two generic evolutionary track morphologies: (1) Models that evolve through the early stages of the AGB but not the TP stage, illustrated by the long-dashed curve in Fig. 1. The evolutionary tracks for these models peel away from the lower AGB before the TP phase as the envelope is consumed from below by nuclear burning and, probably, from above by stellar winds. They are referred to as post-Early AGB (P-EAGB) models (Brocato *et al.* 1990). (2) The least massive of the EHB models never develop extensive outer convection zones and stay at high $T_{eff}(\gtrsim 20,000 K)$ throughout evolution, shown as the short-dashed curve in the diagram. Such models are known as "AGB-Manqué" (failed AGB) sequences (see GR90). Both classes of EHB star may undergo a small number of thermal pulses away from the giant branch. Note also that the range in temperature of the EHB depends on composition. For [Fe/H] < 0, EHB models have $T_{eff} > 10,000 K$ throughout the core helium burning phase, while for [Fe/H] > 0, even some red HB models (cooler than the RR Lyræ instability strip) do not have sufficient mass to reach the upper AGB. In Galactic globular clusters the EHB corresponds to the hottest part of the observed blue tail of the HB (Fusi Pecci *et al.* 1992). The bluest of the EHB models correspond to the field sdB and sdOB stars. Heber (1987) defines these types as stars with $\log g > 5$ and $20,000 K < T_{eff} < 30,000 K$, $T_{eff} > 30,000 K$ respectively. These stars are sometimes referred to by observers as "Extended" HB objects.

The HB lifetimes of the EHB models are $\sim 120 - 150$ Myr, with typical luminosities $10 - 100 L_\odot$. The P-EAGB lifetime is a few hundred thousand years at luminosities of order $1000 L_\odot$. The AGB-Manqué phase of evolution lasts $20 - 40$ Myr at luminosities $\sim 100 - 1000 L_\odot$ and with $T_{eff} \approx 30,000 K$. As will be emphasized in this paper (see also Castellani, Limongi and Tornambè 1992) EHB objects, if present in an old stellar population, will be the largest contributors to the UV flux because of their longevity at high temperature.

GR90 presented a detailed study of the stars that might give rise to the UV excess phenomenon. They investigated the populations of hot stars that would be produced in a metal-rich environment by making simple assumptions about the change in red-giant mass loss with metallicity and about the helium enhancement $\Delta Y/\Delta Z$. They did not study the properties of evolving helium-burning models in detail, but rather concentrated on the effect on the HB mass distribution of their various assumptions. They concluded that the UV-output of post-AGB stars was not sufficiently great to produce the strongest observed UV upturns, and suggested that the hot HB, P-EAGB, and AGB-Manqué stars were probably responsible. They stressed the importance of the stellar population (rather than the intrinsic properties of stellar models) in determining the UV spectrum of evolved stars.

This work presents models necessary for a systematic study of the vacuum UV flux produced by populations of evolved low mass models. Previous work on this subject has estimated the UV radiation from hot stars by using the "Fuel Consumption Theorem" of Renzini & Buzzoni (1986; see also Tinsley 1980). The theorem states that the integrated bolometric luminosity emitted in post-main-sequence stages of evolution is proportional to the fuel consumed in that stage. To avoid the problem of defining the total fuel supply available to post-HB models (*cf.* Castellani, Limongi & Tornambè 1992), as well as to enable the use of bolometric corrections for any given UV waveband, we approach this problem instead by direct calculation. Our grid of tracks is finely spaced at the low-mass end of the HB and fills the region of the HR diagram between the white dwarf cooling sequence for $M_{WD} \sim 0.5 M_\odot$ ($T_{\rm eff} \lesssim 100,000 K$, $\log L/L_\odot < 3.2$) and the AGB. The extensive set of calculations we describe below will be used to produce quantitative estimates of the radiation at short wavelengths ($900 \lesssim \lambda \lesssim 3300$ Å) in later papers in this series.

An important source of uncertainty in the consideration of the high metallicity models arises from the unknown envelope helium abundance. Models with high $Y$ and high $Z$ have the additional benefit that their total mass at fixed age is smaller (GR90), thus reducing the amount of mass loss on the RGB required to produce models on the EHB. Horch, Demarque and Pinnsoneault (1992) have recently found that high metallicity low mass HB stars produce significantly larger amounts of UV radiation than their metal-poor counterparts of similar total mass. We will demonstrate that this result arises mainly from their adopted helium abundances rather than the metallicity. The helium abundance of super-metal-rich stellar populations is, however, unconstrained by observation at present. Models of Galactic chemical evolution (Brocato *et al.* 1990) find $\Delta Y/\Delta Z$ values $\lesssim 1$. Observations have mainly been motivated by obtaining the primordial helium and have concentrated on systems more metal poor than the sun (Skillman 1993; Pagel *et al.* 1992; Peimbert, Torres-Peimbert, & Ruiz 1992). We have therefore investigated the effects of varying $Y$ at high metallicity by computing sequences assuming $\Delta Y/\Delta Z = 0$ (the most conservative assumption), and $\Delta Y/\Delta Z = 3$ or 4.

## 2    *Physical Parameters and Inputs*

Our model calculations have been performed with the evolution code described in Dorman (1992a). We have used a new semiconvection/partial mixing code for the core (Dorman & Rood 1993) which allows easy computation of the end of HB evolution, which is susceptible to "breathing pulses" (Castellani *et al.* 1985; Renzini and Fusi Pecci 1988). We have calculated evolutionary tracks for eight different compositions, and for initial envelope masses $M^0_{\rm env} > 0.002 - 0.005\, M_\odot$. These range from extreme "AGB-Manqué" behavior to those that eventually reach the thermally-pulsing AGB stage. The total number of models computed is extremely large, with about 20 sequences per composition, and $800 - 5000$ models per sequence. As far as possible, we have generated complete tracks for models with envelopes smaller than that needed to reach the thermal pulsing stage on the giant branch. Most of our computations extend to a point on the white dwarf cooling curve where $L = L_\odot$ or where the total hydrogen content $X < 10^{-4}$. We have also computed a small number of sequences through part of the TP-AGB stage in order to estimate the post-AGB contribution to integrated UV fluxes.

The input physics used included the reaction rates tabulated by Caughlan & Fowler (1988), Los Alamos (Huebner *et al.* 1977) and Alexander (1975; 1981, private communication) opacities with the compositions listed below, and the Krishna-Swamy (1966) scaled-solar T-$\tau$ relation for the exterior boundary condition. The equation of state (EOS) used was the Eggleton, Faulkner and Flannery (1973) scheme, which for most of our models becomes equivalent to an ideal arbitrarily relativistic degenerate EOS. This is formally inadequate for representing the physics of the end of our evolutionary sequences when Coulomb interactions are large, but this deficiency is highly unlikely to cause large quantitative changes in the results, at least until the white dwarf cooling sequence is reached. Mass loss is not incorporated in the models; we will discuss its effect on our results for UV output in Section 3.4.

The parameters used for these computations are summarized in Table 1. Zero Age sequences for the models are shown in Figure 2, and the sequences themselves are illustrated in Figure 3. The first column gives the elemental abundance in terms of the iron abundance ratio [Fe/H] of the opacity tables. The next two columns give the value of $Y_{ZAMS}$ and $Y_{HB}$, which differ because of the dredge-up phase on the first giant branch. The fourth column contains the value of $M^0_c$ for each set of sequences, and the fifth column gives the envelope masses ($M^0_{\rm env}$) adopted in each set. The sixth column gives an estimate of the critical envelope mass $M^{\rm TP}_{\rm env}$ required to reach the TP-AGB; models with lower $M^0_{\rm env}$ are EHB sequences by definition. The final column points to the illustration depicting each set of sequences.

There are three sets with metal-poor compositions, [Fe/H] $= -2.26$, [O/Fe] $= 0.50$, [Fe/H] $= -1.48$, [O/Fe] $= 0.63$, (the latter choice being taken from the abundances used by VandenBerg 1992 and Dorman 1992a for which opacity tables were available), and [Fe/H] $= -0.47$, [O/Fe] $= 0.23$. The metal rich models have abundances [Fe/H] $= 0$ ($Z = 0.0169$), $Z = 0.04$ and $Z = 0.06$. The solar metallicity sequences were computed with $Y_{HB} \approx 0.29$. For the more metal-rich sequences, we computed one set with $Y_{HB} = 0.29$, the value obtained by assuming $Y_{ZAMS} = 0.27$ and $\Delta Y/\Delta Z = 0$ for [Fe/H] $> 0$, and including the effects of the first dredge-up. These models have [Fe/H] = 0.39 for $Z = 0.04$ and [Fe/H] = 0.58 for $Z = 0.06$. A second, "enhanced helium" set of sequences was computed based on different assumptions for $\Delta Y/\Delta Z$, taking $\Delta Y/\Delta Z = 3$ for $Z = 0.04$ ($Y_{ZAMS} = 0.34$, [Fe/H] = 0.43), and $\Delta Y/\Delta Z = 4$ for $Z = 0.06$ ($Y_{ZAMS} = 0.45$, [Fe/H] = 0.71).

For all of our chosen compositions, the core masses at the onset of stable core helium burning, denoted $M_c^0$, were determined from the cores of RGB models evolved to the helium flash, with ages between 10 and 15 Gyr. For the metal-poor sequences, the values for $M_c^0$ were taken from computations by VandenBerg (1992) of evolutionary tracks for globular cluster stars, while the core masses for the metal-rich sequences were computed as part of this study. Note that variations in the adopted age do not strongly affect the mass of stars at the tip of the RGB ($\lesssim 0.05\,M_\odot$), in the sense that the effect of age is far smaller than the amount of mass loss required to populate the EHB. The effect of age on $M_c^0$ is practically insignificant in stellar populations older than about 2-3 Gyr.

## 3 Results

### 3.1 Zero-Age Sequences

It is useful before considering the evolutionary sequences themselves to discuss the Zero Age HB sequences, as an aid in the understanding of the factors influencing the energy production during evolution. First, we recall that the ZAHB itself is a sequence of models with decreasing hydrogen shell strength and, to a very good approximation, a fixed core luminosity. Toward its hot end the envelopes of the models are homologous but have decreasing mass (see Dorman 1992b for a discussion). Since the hydrogen is too cool to burn, the envelope acts simply as a "shade" of steadily decreasing thickness. This shade ceases to be inert after core exhaustion. Apart from $M_{env}^0$, the properties of the models are determined by $M_c^0$, $Y$, and $Z$. The metallicity affects both the opacity of the stellar envelope and the rate of CNO burning in the hydrogen burning shell. For $[Fe/H] \gtrsim -1$, the first of these dominates and tends to reduce the energy production $\epsilon_H$ of the hydrogen shell. Increasing $M_c^0$ increases the central temperature and therefore the core helium burning luminosity, while increasing $Y$ increases the hydrogen shell burning temperature and thus $\epsilon_H$. Hence the most metal rich models are fainter because computations of the earlier red-giant evolutionary stages yield smaller $M_c^0$ and because the envelope luminosity is lower. Additionally, as discussed later, the helium burning luminosity is reduced by the higher opacity in the helium-rich core.

Figure 2a shows the ZAHBs on theoretical HR diagram for metallicities up to $Z = Z_\odot$, and implicitly illustrates the relation between luminosity and metallicity. The fifth column of Table I indicates that the lowest mass envelope computed is approximately the same in each curve ($\sim 3 \times 10^{-3} M_\odot$), and it can be readily seen that the temperature of the hottest model in each ZAHB decreases slightly with $Z^1$. At the blue extreme of the ZAHBs, the surface temperatures of the most metal-poor models extend as high as about 30,000K, compared to $\sim 22,000$K for the most metal-rich.

Figure 2b shows the comparison between the two sets of sequences at high metallicity with very different envelope helium abundances. This figure illustrates two separate points. First, the helium rich sequences are much brighter, but their temperature distribution as a function of mass $T_{eff}(M_{env}^0)$ is not strongly affected. Second, the bolometric luminosity at the end of the ZAHB is determined by $M_c^0$, which is significantly smaller for the higher $Y$ sequences (compare Figure 4 of Sweigart & Gross 1976).

From the ZAHB plotted here one can infer that a super-solar metallicity HB star will likely be either very red or very blue. At intermediate temperatures (e.g. $3.7 < \log T_{eff} < 4.1$ for $Z = 0.06$) the mass points are widely spaced (i.e., $d \log T_{eff_{ZAHB}} / d M_{ZAHB}$ is large). Even after allowing for the effects of evolution, this intermediate temperature range will still be under-represented compared with a more metal poor population with similar mass distribution. The stars which begin their HB lifetime at these temperatures will evolve fairly rapidly toward the blue. The early models of HB evolutionary tracks in this range of $T_{eff}$ follow the ZAHB essentially because both comprise sequences of models with decreasing shell brightness. One might anticipate long blueward loops during the evolution (see below and Sweigart & Gross [1976]) for models which lie where the distribution of mass points along the ZAHB is sparse.

---

[1] *This feature is not caused by the variations in $M_c^0$ with metallicity; models with similar $M_c^0$ to the most metal poor models but with solar composition are located on the ZAHB at almost the same temperature – albeit with higher luminosity – as those with the "solar" core mass value.*

## 3.2 Evolutionary Models

Figure 3 (a) through (h) shows the evolutionary tracks. The rate of evolution along the sequences is indicated by filled circles at each 10 Myr through the HB lifetime. We have placed a cross at the point of core helium exhaustion, and subsequent evolution is denoted by open circles at intervals of 2 Myr. To illustrate the late rapid evolutionary stages, filled triangles are placed at 100,000 yr intervals after the time when $L$ reaches $1000\,L_\odot$. Tabulations of the sequences will be available in machine readable format. The tables themselves are described at the end of this paper, where instructions on how to acquire them can also be found.

The features of post-HB evolution are determined by two major factors. The first of these is the delicate interplay between three different sources of luminosity — hydrogen- and helium- shell burning, together with gravitational energy release (referred to below as $L_H$, $L_{He}$, and $L_g$). The second is whether an outer convective zone develops. The diversity of track morphology arises because the similar core evolution is seen through the "filter" of the envelope. Our models indicate that in low mass stars, the formation of an extended giant envelope is directly related to growth in $L_H$ (the other important factor being the steep molecular weight gradient at the edge of the core: see Dorman 1990; Fujimoto & Iben 1991). The strength of the shell itself is regulated by the envelope mass: if $L_H$ remains very small because $M_{env}^0$ is small, the star cannot become a cool giant.

After central helium exhaustion the energy from the triple-$\alpha$ process dwindles until the core regions have contracted enough to ignite a He-shell burning source. This shell-heating phase is quite rapid, and $L_g$ arising from the central contraction is of the same order of magnitude as $L_H$ while $L_{He} \approx 0$. As the evolution proceeds further, however, the helium shell brightens, and by halting and reversing the contraction of the layers above it, it brings a reversal to the rising hydrogen shell and surface luminosity. $L_{He}$ then grows fairly rapidly and eventually reaches nuclear equilibrium, after which its evolution occurs on a nuclear timescale and both $L_H$ and $L$ begin to grow again. Once the helium-burning shell is fully established, the stable He shell burning phase follows, lasting about $\sim 20$ Myr. The behavior of the core region (i.e., $\{M_r : X(M_r) = 0\}$) is similar for all envelope masses.

The behavior is easiest to understand in the extremes. For stars with sufficiently massive envelopes, the growing hydrogen burning shell expands the outer layers to a giant-like configuration during the post-central-exhaustion phase. This leads to a normal AGB evolution. If $M_{env}$ is sufficiently small, however, $L_H$ does not respond to the contraction after central He exhaustion. Without the "push" of an increasing $L_H$ the outer layers do not expand and $T_{eff}$ remains very high. This is AGB-Manqué behavior (GR90). For an intermediate range of $M_{env}$ the brightening of the hydrogen shell forces the model toward the Hayashi line. As hydrogen burning reduces $M_{env}$ the growth of $L_H$ peters out, and the model leaves the lower AGB. Model sequences with this behavior are the post-Early AGB stars (Brocato *et al.* 1990).

Figure 4 illustrates the evolution of three archetypical sequences each for two different metallicities ([Fe/H] = 0 – Fig. 4a and [Fe/H] = $-1.48$ – Fig. 4b). Tables 2 and 3 (a) through (c) list some of the important stellar parameters along the evolutionary tracks depicted in this figure. We have tabulated values of $t_6$, the age in Myr since the ZAHB point, the central helium abundance $Y_c$, $\log L/L_\odot$, $\log T_{eff}$, the surface gravity $\log g_s$, $M_c$ (defined here by the mass at the peak of the hydrogen burning energy curve), and the central conditions $\log T_c$, $\log \rho_c$. These tables give points along the evolutionary tracks at fixed values of $Y_c$ before core helium exhaustion, and at equal intervals along the track thereafter. Although we tabulate values in several of the tracks down to a point of relatively low luminosity on the cooling curve, we stress that the evolutionary tracks here are unreliable for two reasons: (1) our equation of state is inadequate, and (2) our program does not allow the hydrogen shell to burn the outer 0.0001 $M_\odot$.

In both of the AGB tracks in the panels, the evolution stays close to the Hayashi line both during the onset of equilibrium He-shell burning and the later He-shell flash events. The changes in the interior structure resulting from the motion and the growth of the shell sources are masked by the presence of the convective exterior. In the post-exhaustion phase, $L$ and $L_H$ drop as the helium-burning shell becomes fully established. $T_{eff}$ hardly changes, as contraction of the envelope with decreasing $L_H$ merely reduces the depth of the convection zone. After the helium burning shell attains nuclear equilibrium, the evolution enters a slow red phase, sometimes referred to as the "Early AGB phase (E-AGB)" to distinguish it from the later TP-AGB phase. The model proceeds up the AGB until it reaches the thermally-pulsing regime, during which quiescent hydrogen burning at high luminosity is interrupted by violent helium shell flash events that occur on a thermal timescale at intervals of $\sim 10^5$ yr (see Wood & Zarro 1981).

Examples of P-EAGB model sequences are given by the dashed curves in Figure 4. In the upper panel, during the post-core exhaustion phase the evolution reaches a point close to the Hayashi line. As the growth of the hydrogen shell reverses, the outer layers contract with it. The helium-burning shell reaches thermal equilibrium close to the blueward "nose" in this and similar sequences, and the subsequent evolution is redward to the AGB on a nuclear timescale. This part of the track is the E-AGB phase, and a significant part of the track may have $T_{\rm eff}$ significantly greater than $T_{\rm AGB}$. In the corresponding metal-poor track (Fig. 4b), the model is still at high $T_{\rm eff}$ at core exhaustion, and evolves rapidly toward the AGB as $L_{\rm H}$ grows. However, when $L_{\rm He}$ stabilizes, the star is still strongly radiative - note the small feature in the sequence at $T_{\rm eff} \approx 4.25$, $\log L/L_\odot \approx 2$ which indicates the beginning of the E-AGB phase. The other obvious difference between the dashed sequences in the upper and lower panels is the complex (and rapid) loops which represent helium shell flashes occuring after the model has left the giant branch. These are more extreme examples of the changes in surface conditions that can result if the thermal instability in the interior is not masked by an outer convection zone.

The solid curves in Figure 4 are examples of AGB-Manqué tracks. In contrast to the extensive redward evolutionary paths of the models just discussed, the contraction phase is marked by a slight dip in the luminosity as $L_{\rm He}$ diminishes. Since there is no significant hydrogen burning the following evolution is controlled by the growth of the helium burning shell — rather like the core helium burning phase itself – with little or no tendency toward the red. The subsequent evolution thus consists of a "slow blue phase" (see HDP) which is the counterpart to the Early AGB evolutionary phase of more massive model sequences. Eventually as the contraction proceeds further, (i.e., when $\log L/L_\odot \gtrsim 2$), the H-shell at last provides a fair fraction of the luminosity. As it consumes the remaining outer layers, the evolution turns toward very high $T_{\rm eff}$. This phase is not to be confused with the helium-burning main sequence evolution. It is a helium shell-burning phenomenon, and the brighter AGB-Manqué sequences rely on some $L_{\rm H}$ for a fraction of their luminosity. These sequences produce copious UV output throughout the late stages of evolution, and they are distinguished from each other only by the size of the hydrogen burning contribution to the luminosity at late times. Their longevity allows the possibility of strong observational tests for their existence. The ratio of stars that lie in the AGB-Manqué stage to their EHB precursors is predicted to be 1:5 to 1:6 depending on composition, and a significant population of these should be detectable in star clusters and the Galactic Bulge using space based UV photometry.

We next discuss the upper bound to the masses of EHB stars, as defined in §1. P-AGB stars arise after the TP-AGB when the envelope is exhausted by nuclear burning and by stellar winds. There is clearly a critical size of envelope mass required for reaching this stage, equal to the hydrogen fuel consumed during AGB evolution, plus any mass loss suffered prior to the TP-AGB. The value of $M_{\rm env}$ — and therefore the fuel supply — at the point where the rapid blueward evolution commences is bounded above by stellar structure considerations. If the envelope is too large, i.e., $M_{\rm env} \gtrsim 0.002\,M_\odot$ (decreasing with $L$), the model will not leave the giant branch. The physical processes by which P-AGB evolution begins can be qualitatively understood as follows. In the quiescent H-burning phases, $L = L_{\rm H}$, and the luminosity increases with the core mass. By definition, a model stays on the giant branch as long as it can remain in equilibrium with a cool convective envelope. The steeper temperature gradient arising from the brighter burning shell in more luminous models allows stars with smaller envelope masses to exist with exterior convection zones. Eventually, however, this is not possible: the outer layers must contract so that the density in the burning shell is high enough to keep it in equilibrium, and the star evolves off the giant branch. This process will be assisted by mass loss (*cf*. Schönberner 1983), which increases with luminosity and with decreasing temperature (see section 3.4).

The duration of the post-AGB evolution is related to the luminosity at which the star crosses the HR diagram and the remaining fuel supply. Figure 5 shows the evolution of solar metallicity sequences for total masses $M_{\rm HB} = 0.50, 0.51, 0.52$ and $0.53\,M_\odot$ ($M_{\rm env}^0 = 0.041$ to $0.071\,M_\odot$). For the two fainter sequences, open circles mark 0.25 Myr intervals. These are examples of P-EAGB stars, and their evolutionary timescale in this phase is of order 1 Myr. For the two brighter sequences, triangles mark 10,000 year evolutionary intervals on the part of the track where $\log L/L_\odot > 3$. The most luminous of these sequences is an example of low-mass P-AGB evolution according to our definition, since it enters the thermally pulsing stage while still on the AGB. Its rate of evolution is very high compared to the others. The $0.52\,M_\odot$ track has two thermal pulses which occur after the star has started to cross the HR diagram, and represents a model very close to the transition between classic P-AGB behavior and the Post-Early AGB morphology. The difference in evolutionary rate arises, in addition to the change in luminosity and hydrogen fuel supply, from the fact that post-Early AGB stars are partially supported by He-burning as they evolve to higher temperatures (Castellani, Limongi & Tornambè 1992). Since $M_c^0([Fe/H] = 0) = 0.469\,M_\odot$, we find that EHB behavior is bounded by $M_{env}^0 \sim 0.05 M_\odot$.

The sixth column in Table 1 lists the $M_{env}^0$ values for the most massive sequence that does not begin thermally pulsing on the AGB, and this represents the boundary between the models that evolve into P-AGB sequences and those that do not. Suprisingly, the total masses of the models in this column are the same for all compositions with $[Fe/H] \leq 0$. For the most metal-rich sequences, the models with $M_{env}^0 < M_{env}^{TP}$ as listed in the table do not undergo thermal pulses at all, as we will discuss briefly in a later section.

Schönberner (1979, 1983) computed P-AGB sequences by evolving models of several masses through the HB phase toward the upper AGB. His HB evolution differed from ours as partial mixing (semiconvection) was not included: the effect of this is to lengthen the AGB lifetime at the expense of the HB evolution. He then applied high mass loss rates to the models at selected luminosities, so that each model began to evolve as a Post-AGB sequence from that point. The masses of the remnants at each selected value of $L$ have been used to derive a low mass $L$-$M_c$ relationship, and the timescales for evolution to the cooling track have also been used to estimate the UV output from low mass P-AGB stars, as well as in studies of planetary nebulæ. Our sequences differ in the core masses found for evolving models at his chosen luminosities, which we assume arises because more time is spent at higher luminosity. For example, a model sequence almost identical to his 0.8 $M_\odot$ track reached the TP-AGB stage at $L = 2000 L_\odot$ and $M_c = 0.535\,M_\odot$, and had $M_c = 0.526\,M_\odot$ at $L = 1400 L_\odot$, compared to his sequence that had $M_c = 0.545\,M_\odot$ at similar luminosity. Despite the difference in the core mass, our sequence with $M \approx M_c^{P-AGB} = 0.53\,M_\odot$ has very similar transition luminosity to Schönberner's 0.565 $M_\odot$ sequence. This Schönberner sequence crosses from $T_{eff} = 5000\,K$ to $100,000\,K$ in about 16,000 yr, with $\log L/L_\odot = 3.60$ as compared to about 26,000 yr, $\log L/L_\odot = 3.55$ for our model.

One can determine the potential contribution of a given type of star to the UV light of a stellar population by integrating the UV luminosity over the individual evolutionary tracks. A typical result of such a calculation is shown in Fig. 6 where the UV energy output from the evolutionary sequences for $Z_\odot$ is shown as a function of $M_{env}^0$. We have chosen a bandpass of width 150Å centered close to 1500Å for this particular figure, and the energy output is given in units of $10^{48}$erg Å$^{-1}$ per star. The luminosity in the band has been estimated using bolometric corrections derived from the Kurucz (1991) model stellar atmospheres for $T_{eff} < 50,000 K$, and those of Clegg & Middlemass (1987) for higher surface temperatures. The points corresponding to the models illustrated in Fig. 5 are labelled by their total mass. For the 0.53 $M_\odot$ model, the P-AGB contribution is almost insignificant compared to the output during the HB phase. Models that enter the P-AGB stage with still larger masses have higher luminosity and smaller fuel supply, and thus more rapid evolution and smaller UV energy output. Apart from the P-AGB contribution, at this metallicity only the EHB stars and their progeny emit significant UV radiation. Note that the UV output decreases toward the point representing the helium main-sequence model, a circumstance that arises partly because of larger bolometric corrections but also from the decreasing luminosity. The value of $M_{env}^0$ for which the flux is greatest is a weak function of the bandpass chosen, and for all but the shortest wavelengths it lies close to $M_{env}^0 = 0.02 M_\odot$. We will discuss the variation of these curves with bandpass and with stellar composition in much greater detail in Paper II.

### 3.3 Evolution of Models with High Metallicity

The figures 3a – 3f show a gradual change in morphology of the EHB sequences during the post-HB stages of evolution. The most obvious difference is the relative paucity of tracks crossing the region of intermediate temperatures ($\log T_{AGB} \lesssim \log T_{eff} \lesssim 4$) among the metal-rich sequences as compared to the metal poor sequences. This can be traced to the ZAHB mass distribution, in which which is predominantly red for the entire evolution, while the models which are blue on the ZAHB evolve into AGB-Manqué objects. In contrast, for low metallicity (see Figs 3a and 3b), stars with envelopes which have ample mass for the star to reach the AGB can be quite hot during the HB evolution, and since outer envelopes do not become convective until the star has become very bright, they evolve redward for most of the E-AGB phase.

Models predict that the HB stars of metal-rich populations will differ in subtle ways from those found in globular clusters. Differences in the composition causes changes in the interiors of the models that affect their luminosities and their rates of evolution. For metal-poor compositions, the free-free contribution to the core opacity is generally about 25% (Dorman & Rood 1993) and is largely due to carbon produced during the helium core flash. For [Fe/H] > 0 however, the core material at the ZAHB has up to twice the carbon abundance in addition to a contribution from the iron-peak elements. This abundance difference, together with the smaller core masses predicted by RGB evolution, produces smaller central temperatures and helium-burning luminosities. As a result, the central conditions that obtain in lower metallicity models at the ZAHB are not encountered until 10–20 Myr into the evolution. The initial size of the convective core is smaller and the core expansion (overshooting) phase is prolonged. The circumstances that induce a partially mixed (semiconvective) region surrounding the convective core do not arise until the central helium abundance $Y_c \approx 0.60$, compared with $Y_c > 0.70$ found for metal-poor models. The convective core size grows to about $0.14 - 0.15 \, M_\odot$, and the greatest extent of the partial mixing zone is at about $0.22$–$0.24 \, M_\odot$, as compared to $0.17 \, M_\odot$ for the convective core and $0.27$–$0.29 \, M_\odot$ for the mixing zone in the most metal-poor sequences. The total supply of helium accessible to the core burning source is thus reduced. Nevertheless, the core luminosity is also lower, and the time taken to exhaust helium in the core is about 20% greater for $Z = 0.06$ than for $[Fe/H] = -2.2 (Z = 0.0001)$. The core mass also affects the rate of HB evolution through its effect on the central temperature: since $M_c$ increases with time if $L_H$ is large, the more massive models with stronger shell burning will have lifetimes similar to those of less metal-rich models. As a result the metallicity effect on HB timescales is obvious only for the EHB models. For the post-HB evolution, the smaller extent of the mixing region on the HB implies that the amount of helium left over at central helium exhaustion is larger. The slow blue phase is therefore somewhat longer, lasting up to 30 Myr in the most metal-rich sequence as opposed to about 20 Myr at low metallicity. However, their luminosity is lower than the more metal rich objects, and the UV energy output of models in this phase will not be very different. The helium abundance of the models illustrated in Figs 3e and 3f is, however, almost certainly underestimated, while our computations with $Z = 0.06, Y_{\mathrm{ZAMS}} = 0.45$, (assuming $\Delta Y/\Delta Z = 4$) represent the other extreme.

The tracks computed with enhanced helium abundance show important differences in the HB phase and in the rapid evolutionary phase following He-exhaustion. As noted above, the essential difference resulting from enhanced helium abundance is a major increase in the hydrogen burning luminosity, all other parameters being fixed. Since $Y$ does not strongly affect the opacity, it does not greatly affect the ZAHB mass distribution, and ZAHB models lie on the red HB for all but the smallest $M_{\mathrm{env}}^0$. The blueward loops in the HB evolutionary sequences arise from a contraction in the envelope caused by a decrease in $L_H$ with time. This is, in turn, results from expansion of the core region in the early part of the evolution. The effect is most pronounced when both $Y$ and $Z$ are large: in the subsequent evolution off the ZAHB, $L_H$ decreases much more dramatically than with lower $Y$, and the blueward loop extends to much higher temperatures. These extended loops move the models into the far-UV "window" for a greater part of their core He-burning lifetime. The core begins to contract long before core exhaustion and takes the evolution back toward the red. Since the core mass is significantly larger than on the ZAHB, the hydrogen burning shell brightens rapidly to a higher luminosity. The result of both the helium and hydrogen shells increasing together is a wide loop in the HR diagram (Sweigart & Gross 1976).

The post-HB evolutionary morphology is also modified as a result of the enhanced hydrogen burning. The envelope is consumed more rapidly, and the critical mass $M_{\mathrm{env}}^{\mathrm{TP}}$ thus increases. In the AGB-Manqué-type objects, $L_H$ is suppressed by the small envelope mass throughout evolution, just as in the lower-$Y$ sequences. However, at core exhaustion, the hydrogen shell responds more strongly than is the case with lower $Y$. Since the outer layers are more easily consumed before the model reaches the AGB, a larger range of masses evolves with the AGB-Manqué-type morphology. Note that the much greater lifetime of the sequences in Fig. 3g compared to Fig. 3e is mainly due to the difference in $M_c^0$. Thus the UV output from the high-$Y$ sequences is significant for a greater range in $M_{\mathrm{env}}^0$ than in their low-$Y$ counterparts. These enhanced-helium sequences also demonstrate that the post-AGB relationship between $L$ and $M_c$ depends on $Y$. The enhanced fuel consumption produces a core mass $M_c > 0.6 M_\odot$ at $\log L \sim 3.3$, a luminosity very much smaller than that expected from the Paczyński (1971) relation, for models with $M_{\mathrm{env}}^0 \approx M_{\mathrm{env}}^{\mathrm{TP}}$. The critical mass for evolution to the TP-AGB is also larger (see Table 1).

HDP recently studied the evolution of similar objects. They computed models for $Z = 0.02, 0.04$, and $0.06$ with two values for $Y$ for each metallicity, both assuming relatively large values of $\Delta Y/\Delta Z$. Their models indicated that at higher metallicities, a larger range of masses becomes AGB-Manqué stars, and this was especially true of their higher helium abundances. They pointed out that if helium abundance increases with $Z$ as steeply as $\Delta Y/\Delta Z = 4$ for [Fe/H] > 0, then HB models with the range of masses normally inferred from the HB — rather than those with enhanced degrees of mass loss — could be responsible for excess UV flux in elliptical galaxies. Some of our computations were conducted with very similar choices of the abundance to HDP's. We are able to reproduce some of their evolutionary sequences, but for the set with $Y \sim 0.45$, $Z = 0.06$, our sequences are very different. None of our models possess a stage in which the hydrogen in the envelope is exhausted before helium shell burning takes place. In addition, we find that the transition mass ($M_{tr}$), below which the models become AGB-Manqué stars, is in the range $0.55 - 0.575 \, M_\odot$, rather than the value of $0.80 M_\odot$ found by HDP. For $Z = 0.04$ our sequences with $Y = 0.36$ have $M_{tr}$ at about $0.52 \, M_\odot$, compared to their $M_{tr} = 0.62 \, M_\odot$ for models with $Y = 0.37$. However, comparison with our lower $Y$, high-$Z$ sequences shows that it is the envelope helium abundance, rather than the metallicity, that is responsible for the larger range of masses that may produce UV radiation. HDP's argument that higher CNO abundance in high-metallicity models gives rise to more rapid hydrogen burning neglects the effect of opacity in regulating $L_H$ (see Dorman 1992b).

One very noticeable difference between the metal-rich and metal-poor evolutionary sequences is in shell flash phenomena. The lower mass models with $Z > Z_\odot$ experience the shell flashes only on the AGB, whereas models of similar $M_{env}^0$ but lower metallicity sometimes undergo flashes during hotter phases. Some of our sequences also exhibit a hydrogen shell flash phenomenon close to the cooling track, triggered at the point where the contracting model heats the remaining hydrogen (usually $X \lesssim 10^{-4} \, M_\odot$). These events are particularly noticeable in Figs 3c and 3d.

It is clear that we expect composition differences to affect the range of masses that undergo thermal pulses on the AGB itself. The obvious tendency for more metal-rich models to possess exterior convection zones makes this phenomenon more likely to take place on the AGB rather than at some intermediate temperature. However, for the most metal poor sequences, every sequence undergoes a thermal pulse at some point on the track, whereas for metal rich sequences only the more massive models do so[2]. This has also been noted by Castellani & Tornambè (1991), who suggest that the opacity must be responsible in suppressing the shell flash phenomenon. The few numerical experiments we have conducted on this issue concur with that suggestion. For our purposes here we note that neither kind of high temperature shell flash is, if they do occur in nature, is of sufficient duration to affect the UV output from a population. This is especially true for the hydrogen shell flash phenomena (also found by Caloi 1989; Castellani & Tornambè 1991, and HDP), which cause rapid brightening for short periods ($\sim 1000$ yr) at very high temperatures.

### 3.4  The Effects of Mass Loss

An important omission from the models in this study is that of mass loss during the HB and AGB phases. The actual evolution toward and beyond the TP-AGB phase is almost certain to be seriously affected by the stellar winds observed in cool giants (see Dupree 1986 for a review) and inferred from the HB morphology (Iben and Rood 1970; Rood 1973; Lee, Demarque & Zinn 1990). Stars on the AGB sequence are sufficiently similar in their exterior properties that they should be subject to the same slow mass loss processes that occur on the RGB. We exclude the unlikely circumstance that the bulk of RGB mass loss takes place at the helium core flash. The upper AGB evolution could conceivably have greater mass loss, related to physical processes such as thermal pulses not found on the RGB, in addition to the mechanism that is responsible for the formation of planetary nebulæ. Thus even for luminosities less than the helium core flash luminosity ($L \sim 1800 - 2500 L_\odot$) each star has two phases in which a few tenths of a solar mass may be lost. The 'slow' stellar wind is often parametrized by the mass loss formula deduced by Reimers (1975), i.e.,

$$\dot{M} = -4 \times 10^{-13} \eta \frac{L}{gR} M_\odot \, \mathrm{yr}^{-1},$$

---

[2] *There is a numerical aspect to shell flashes found by our code that are set off at high temperature; we have found, for example, that model sequences that run into He shell flashes may in some instances be evolved quiescently to the cooling curve. It is possible that small changes in the composition (from numerical problems in the code arising from mesh optimization procedures etc) in certain interior shells may cause variations in numerical results. However, the composition dependence of the behavior noted in this section is unambiguous.*

(where $L$, $g$ and $R$ are expressed in solar units), in which $\eta$ is a free parameter whose value is constrained by observation. Renzini (1981b) points out that $\eta \lesssim 0.6$ for red giant stars, as larger values would inhibit core helium burning and force the models directly to the cooling track. Horizontal branch morphology generally requires $\eta \sim 0.25$–$0.5$ for Galactic globular clusters. Further, unless a similar value is maintained on the AGB one would find many AGB stars in globular clusters significantly brighter than the tip of the red giant branch, contrary to observation.

Let us assume that mass-losing stars readjust on a thermal timescale, so that the star has the structure of a model from a constant-mass sequence of the same envelope mass at the stage of evolution appropriate to the state of the core. We consider the possibilities for mass loss (as parametrized by the Reimers formula) during helium-burning evolution. The tightly-bound AGB-Manqué stars will lose very little mass (a few $\times 10^{-4} M_\odot$) during their evolution, given $\eta \lesssim 1$. Their evolution is thus adequately represented by constant-mass sequences. For the models with $M_{env}^0 > M_{env}^{TP} + \delta M_{AGB}$, ($\delta M_{AGB}$ here refers to the total amount of mass lost on the AGB during stable He shell burning) the evolutionary tracks for different envelope masses are closely coincident, and thus the evolutionary tracks will be virtually unaltered by mass loss. The UV output will be given in principle by P-AGB evolutionary sequences.

For a range of $M_{env}^0$, the lifetime integrated UV flux may be increased by mass loss. Figure 7 shows an estimate of the cumulative mass loss predicted by the Reimers formula for four of the solar metallicity sequences, in units of $M_\odot/\eta$, as a function of the luminosity (these *post hoc* assessments will be overestimates for models with temperatures and surface gravities which are significantly increased as a result of the mass loss). The shell flashes have been suppressed for the sake of clarity. The start of Early-AGB evolution at $\log L/L_\odot \sim 2.2$ can be seen clearly in the diagram, and the total mass loss increases rapidly beyond $\log L/L_\odot = 3$. The envelope mass can either be reduced enough so that the model enters the AGB-Manqué region of the HR diagram, or simply leaves the AGB early like the constant-mass P-EAGB models. For example, unless $\eta$ is very small, the two smallest masses illustrated in Fig. 7 will not proceed very far toward the giant branch before losing sufficient mass to adopt the AGB-Manqué behavior, while the 0.60 $M_\odot$ model may become a P-EAGB star if $\eta \gtrsim 0.4 - 0.5$.

It seems probable that only the less massive stars can become AGB-Manqué objects as a result of mass loss. The UV contribution from these stars will increase, but to an amount not exceeding the greatest difference between the upper and lower curves in Fig. 6. For the more massive sequences, the lifetime UV output is in the range spanned by the post-HB contribution from the constant mass P-EAGB sequences. In addition since the most rapid mass-losing phases are at high luminosity, which is attained late in the evolution, this output is likely to be close to the lower end of the range. These arguments can be summarized by the statement that the lifetime UV output from a mass-losing model must be less than that from a constant-mass model sequence with the same final (white dwarf) mass. Hence our derived radiation curves such as Figure 6 represent an upper bound to the integrated UV flux from objects of terminal mass $M_c^0 + M_{env}^0$.

## 4   *Summary*

We have computed an extensive grid of horizontal-branch and post-HB models for a large range in composition. These sequences expand the parameter space of advanced stages of stellar evolution that has been explored in detail, in metallicity and in range of masses. We have calculated HB sequences for a number of super metal-rich compositions, to be of use in interpreting observations of hot stars in elliptical galaxies and in the Galactic bulge and field. Our large set of computations for EHB stars will be a component of future studies of the blue tail of the HB of globular clusters. Such studies may provide new insights into the globular cluster populations and the relation to their dynamical properties (see Fusi Pecci *et al.* 1992), because they will focus on a population which is resolved in cluster cores.

We have also explored the low mass end of post-AGB evolution in some detail. We note that if the core masses of HB stars are approximately fixed, then a scatter in the masses attained by the core at the end of the AGB phase is a natural consequence of the variation in hydrogen burning luminosity on the HB. The effect is of order 0.01 $M_\odot$ between models with $M_{ZAHB} = 0.60$ and $0.90\,M_\odot$. We find that the total mass required to reach the TP-AGB is quite insensitive to $Z$. The sequences also indicate that UV-bright stars in globular clusters with $L \approx 1000 - 1800 L_\odot$ like many of those observed (deBoer 1985; Landsman *et al.* 1992), should have masses $\lesssim 0.52\,{\rm M}_\odot$. This luminosity places the stars below the region of thermal pulses, i.e., in the P-EAGB regime. The number counts of such stars relative to the numbers of HB stars is consistent with lifetimes of a few hundred thousand years for these objects (Renzini & Buzzoni 1986), significantly longer than the true P-AGB stars. Recall that the faintest model sequence computed by Schönberner (1983) is a hydrogen-burning P-EAGB star by our definition: it does not reach the TP-AGB stage because of the assumed mass loss (see also Castellani & Tornambè 1991).

If the mass loss on the first red giant branch is such that the EHB is significantly populated then the UV light from the EHB itself, AGB-Manqué, and P-EAGB stars will vastly exceed that from P-AGB stars in an old stellar population. The apparent correlation of the UVX with metallicity (or more accurately, with the absorption line strength of magnesium) provides the only clue so far uncovered as to why some galaxies might produce more UV-bright stars.

The general features of the evolutionary sequences as a function of $\rm M_{env}^0$ vary with the composition of the envelope. As pointed out by Caloi (1989), for any given choice of metallicity and $Y$ each class of sequences (AGB-Manqué, P-EAGB, etc) is present for some range of masses (*cf.* Ciardullo & Demarque 1978). It is thus possible that an old population of any composition could have significant UV light: whether such a population actually does so will depend on the mass distribution of stars arriving on the HB. The UV output of HB sequences is, of course, primarily determined by $\rm M_{env}^0$ for models of fixed composition. The UV energy integrated along an evolutionary track is approximately proportional to the flux from a population of stars of a single mass, arriving on the HB at a constant rate. This arrival rate of stars on the HB can be predicted from the number of stars leaving the main sequence or the population of the upper part of the giant branch. The flux from a stellar population can thus be derived by assuming a distribution of HB stars in $\rm M_{env}^0$, and a knowledge of the lifetime-integrated UV energy as a function of $\rm M_{env}^0$, $E_\lambda(\rm M_{env}^0)$, for a bandpass or filter represented by $\lambda$. Simple arguments suggest that the function $E_\lambda(\rm M_{env}^0)$ derived from constant mass evolution sequences can be used to find reliable estimates for the UV radiation from mass-losing stars. The greatest UV flux is produced by stars which lost their mass prior to arrival on the HB.

The UV energy radiated during the HB lifetime is very significant for blue HB models. For the most metal-poor compositions, the more massive of these stars eventually reach the P-AGB stage, and their post-HB UV flux will not contribute greatly to their total output. The EHB stars all produce large post-HB flux; this contribution rises as the mass is decreased through the post-Early AGB to the AGB-Manqué mass range, in which the greatest integrated energies are found. $E_\lambda(\rm M_{env}^0)$ is found to peak for some $\rm M_{env}^0 \gtrsim 0.02\,{\rm M}_\odot$, the exact point depending on the UV waveband.

For metallicities above solar, the conclusions depend on the helium abundance. For one extreme assumption about $Y$, $\Delta Y/\Delta Z = 0$, the EHB model sequences are fainter and longer-lived than more metal poor sequences, as well as being somewhat cooler, and only a small mass range produces any UV radiation. This range tends to increase with $Y$ both because more models become hot during their HB existence and because a larger mass is needed to reach the TP-AGB stage. A large value of $Y$ appears to be necessary to make a big difference in this range of masses (see paper II). While the UV output from high $Y$ models is expected to be somewhat larger than from the low $Y$ models, the assumption of very high $Y$ at high $Z$ is not necessary to explain the magnitude of the UV output from a metal-rich population. However, the degree of mass loss required to produce hot stars decreases with $Y$ because (1) the red-giant progenitor stars of fixed age may be significantly less massive if $\Delta Y/\Delta Z$ is large, e.g., by about $0.1\rm M_\odot$ between $Y = 0.27$ and $Y = 0.45$ at $Z = 0.06$, and (2) more massive HB and post-HB objects are UV producers. Our models indicate that for the highest $Y$ and highest $Z$ models, a UV-bright population might be produced with degrees of mass loss that are comparable from that inferred from the globular clusters. On the other hand, for sufficiently low $\Delta Y/\Delta Z$, a metallicity-enhanced mass loss function appears to be the only possibility allowing large UV fluxes from evolved metal-rich stars. It is likely that mass loss does indeed increase with metallicity, at least for [Fe/H] $\geq 0$. One of the most popular current models for mass loss in very cool stars involves radiation pressure on dust grains (see MacGregor & Stencel 1992), which will become more efficient at low surface temperatures and high metal abundance. Very recently, Tripicco, Dorman & Bell (1993) have found that higher mass loss rates than those inferred for globular cluster stars are needed to explain the color of the clump stars in the open cluster M 67, which has solar metallicity.

In sum, comprehending how the UVX phenomenon may be correlated with any parameter requires an understanding of how that parameter affects the mass distribution on the ZAHB. A second ingredient must be a better knowledge of galactic chemical evolution as it affects the helium abundance at high $Z$. The development of observational tests for $Y$ in metal-rich stellar populations (difficult enough in any environment) would be very useful. Whether the UVX or analogous phenomena can be used as a probe for parameters, such as age, of old stellar populations must await a better understanding of these factors.

Progress in understanding the stellar population responsible for the UVX phenomenon requires quantitative comparisons between synthesized observable quantities, in particular the $1500 - V$ color, and the observations themselves. The following papers in this series will use these sequences to provide such estimates, and discuss in detail the integrated UV fluxes predicted by these models. We will also produce synthetic integrated spectra that illustrate variations of the UV flux with the stellar population, and compare these to observation.

*Note concerning the tabulations and their acquisition:* The tabulated data are available in machine-readable form either from the *Astronomical Data Center* (ADC) or the *Centre de Données astronomiques de Strasbourg* (CDS). These data may be obtained in machine-readable form, either on magnetic tape or possibly via transmission over an electronic network. To obtain the data from the ADC, one may complete and submit the form published in the latest issue of the *Astronomical Data Center Bulletin*, or transmit the information on the form to the ADC by telephone [(301)-286-8310] or electronic mail to any of the following addresses: BITnet: TEADC@SCFVM; Internet: teadc@scfvm.gsfc.nasa.gov; NSI-DECnet (formerly SPAN) NSSDCA::ADCREQUEST. Data can be obtained from CDS by submitting the form published in the latest issue of the *Bulletin d'Information du CDS*, or by sending the request via electronic mail to question@simbad.u-strasbg.fr (Internet) or SIMBAD::QUESTION (SPAN).

We would like to thank Pierre Demarque for interesting discussions and Bob Hill and Wayne Landsman for information concerning UV-bright stars. We are also grateful to Wayne Landsman for supplying us with the IDL Astronomy Users' Library and an IDL database of the Kurucz stellar atmospheres. This research was supported by NASA Long Term Space Astrophysics Research Program grant NAGW-2596.

16*Table 1: Model Calculations*

| [Fe/H] [O/Fe] | $Y_{\text{ZAMS}}$ | $Y_{\text{HB}}$ | $M_c^0$ | $M_{\text{env}}^0$ ($M_\odot$) | $M_{\text{env}}^{\text{TP}}$ ($M_\odot$) | Figure |
|---|---|---|---|---|---|---|
| −2.26 0.50 | 0.235 | 0.245 | 0.495 | 0.003, 0.005, 0.010, 0.015, 0.020, 0.025, 0.035, 0.045, 0.065, 0.085, 0.105, 0.145, 0.225, 0.285, 0.405 | 0.025 | 3a |
| −1.48 0.63 | 0.236 | 0.247 | 0.485 | 0.003, 0.005, 0.010, 0.015, 0.020, 0.025, 0.035, 0.045, 0.055, 0.075, 0.105, 0.125, 0.145, 0.165, 0.215, 0.295, 0.415 | 0.035 | 3b |
| −0.47 0.23 | 0.238 | 0.257 | 0.475 | 0.003, 0.005, 0.010, 0.015, 0.020, 0.025, 0.035, 0.045, 0.065, 0.085, 0.125, 0.185, 0.225, 0.305, 0.405 | 0.045 | 3c |
| 0.00 0.00 | 0.270 | 0.288 | 0.469 | 0.002, 0.004, 0.006, 0.011, 0.016, 0.021, 0.026, 0.031, 0.041, 0.051, 0.061, 0.071, 0.081, 0.091, 0.111, 0.151, 0.191, 0.231 0.431, 0.531 | 0.051 | 3d |
| 0.39[a] 0.00 | 0.270 | 0.292 | 0.464 | 0.003, 0.004, 0.006, 0.011, 0.016, 0.021, 0.026, 0.036, 0.046, 0.056, 0.066, 0.076, 0.096, 0.116, 0.136, 0.236, 0.336 | 0.076 | 3e |
| 0.58[b] 0.00 | 0.270 | 0.289 | 0.458 | 0.003, 0.005, 0.007, 0.012, 0.017, 0.022, 0.032, 0.042, 0.052, 0.062, 0.082, 0.102, 0.142, 0.242, 0.342, 0.442 | 0.062 | 3f |
| 0.43[a] 0.00 | 0.340 | 0.356 | 0.454 | 0.003, 0.005, 0.007, 0.011, 0.016, 0.021, 0.026, 0.036, 0.046, 0.056, 0.066, 0.076, 0.086, 0.096, 0.121, 0.246, 0.346, 0.446 | > 0.10 | 3g |
| 0.71[b] 0.00 | 0.450 | 0.459 | 0.434 | 0.006, 0.011, 0.016, 0.021, 0.026, 0.036, 0.046, 0.056, 0.066, 0.076, 0.086, 0.106, 0.116, 0.141, 0.166, 0.216, 0.266, 0.366 | > 0.14 | 3h |

[a] $Z = 0.04$
[b] $Z = 0.06$

TABLE 2

SAMPLE METAL-RICH EVOLUTIONARY SEQUENCES:

$Z = 0.04$, $Y_{\rm HB} = 0.29$, $M_c = 0.464\,{\rm M}_\odot$

| $t_6$ | $Y_c$ | $\log L/L_\odot$ | $\log T_{\rm eff}$ | $\log g_s$ | $M_c$ | $\log T_c$ | $\log \rho_c$ |
|---|---|---|---|---|---|---|---|
| A. $M_{\rm env}^0 = 0.003\,{\rm M}_\odot$ | | | | | | | |
| 1.231 | 0.9184 | 1.0222 | 4.3783 | 5.5513 | 0.4649 | 8.0624 | 4.3622 |
| 2.860 | 0.9000 | 1.0260 | 4.3781 | 5.5465 | 0.4649 | 8.0631 | 4.3626 |
| 13.718 | 0.8000 | 1.0476 | 4.3763 | 5.5178 | 0.4650 | 8.0677 | 4.3624 |
| 27.059 | 0.7000 | 1.0723 | 4.3741 | 5.4842 | 0.4649 | 8.0731 | 4.3589 |
| 42.712 | 0.6000 | 1.0988 | 4.3716 | 5.4479 | 0.4649 | 8.0794 | 4.3520 |
| 60.145 | 0.5000 | 1.1277 | 4.3695 | 5.4104 | 0.4649 | 8.0869 | 4.3460 |
| 77.756 | 0.4000 | 1.1581 | 4.3681 | 5.3746 | 0.4649 | 8.0957 | 4.3442 |
| 94.431 | 0.3000 | 1.1895 | 4.3681 | 5.3430 | 0.4649 | 8.1065 | 4.3490 |
| 109.276 | 0.2000 | 1.2235 | 4.3706 | 5.3190 | 0.4649 | 8.1211 | 4.3652 |
| 124.069 | 0.1000 | 1.2604 | 4.3796 | 5.3181 | 0.4649 | 8.1442 | 4.4083 |
| 132.997 | 0.0500 | 1.2805 | 4.3919 | 5.3471 | 0.4649 | 8.1655 | 4.4594 |
| 137.926 | 0.0250 | 1.2908 | 4.4051 | 5.3898 | 0.4649 | 8.1844 | 4.5127 |
| 140.892 | 0.0100 | 1.3020 | 4.4214 | 5.4441 | 0.4649 | 8.2047 | 4.5809 |
| 142.563 | 0.0010 | 1.3277 | 4.4567 | 5.5591 | 0.4654 | 8.2409 | 4.7355 |
| 142.751 | 0.0001 | 1.3516 | 4.4701 | 5.5890 | 0.4654 | 8.2493 | 4.8047 |
| 142.804 | 0.0000 | 1.3614 | 4.4751 | 5.5991 | 0.4654 | 8.2493 | 4.8343 |
| 144.216 | 0.0000 | 1.4300 | 4.5201 | 5.7107 | 0.4649 | 8.1899 | 5.1725 |
| 148.392 | 0.0000 | 1.6298 | 4.5168 | 5.4977 | 0.4651 | 8.0996 | 5.2775 |
| 162.303 | 0.0000 | 1.8246 | 4.5619 | 5.4831 | 0.4649 | 8.0810 | 5.4716 |
| 171.503 | 0.0000 | 2.0039 | 4.6415 | 5.6224 | 0.4648 | 8.0390 | 5.7106 |
| 175.542 | 0.0000 | 2.0107 | 4.7720 | 6.1375 | 0.4657 | 7.9886 | 5.8722 |
| 176.064 | 0.0000 | 2.0722 | 4.8462 | 6.3726 | 0.4662 | 7.9885 | 5.9114 |
| 176.207 | 0.0000 | 1.8722 | 4.8453 | 6.5689 | 0.4663 | 7.9898 | 5.9264 |
| 176.317 | 0.0000 | 1.6966 | 4.8345 | 6.7015 | 0.4664 | 7.9900 | 5.9374 |
| 176.446 | 0.0000 | 1.5477 | 4.8282 | 6.8252 | 0.4664 | 7.9893 | 5.9493 |
| 176.608 | 0.0000 | 1.3584 | 4.8143 | 6.9587 | 0.4665 | 7.9873 | 5.9628 |
| 176.806 | 0.0000 | 1.1629 | 4.7967 | 7.0841 | 0.4665 | 7.9833 | 5.9777 |
| 177.069 | 0.0000 | 0.9647 | 4.7749 | 7.1948 | 0.4665 | 7.9764 | 5.9947 |
| 177.425 | 0.0000 | 0.7666 | 4.7484 | 7.2875 | 0.4665 | 7.9651 | 6.0140 |
| 177.929 | 0.0000 | 0.5690 | 4.7179 | 7.3627 | 0.4665 | 7.9471 | 6.0359 |
| 178.631 | 0.0000 | 0.3719 | 4.6836 | 7.4224 | 0.4665 | 7.9214 | 6.0586 |
| 179.578 | 0.0000 | 0.1754 | 4.6464 | 7.4705 | 0.4665 | 7.8897 | 6.0797 |
| 180.687 | 0.0000 | −0.0001 | 4.6116 | 7.5066 | 0.4665 | 7.8586 | 6.0963 |
| B. $M_{\rm env}^0 = 0.046\,{\rm M}_\odot$ | | | | | | | |
| 1.033 | 0.9181 | 1.3146 | 4.0097 | 3.8225 | 0.4649 | 8.0647 | 4.3400 |
| 2.626 | 0.9000 | 1.3273 | 4.0029 | 3.7827 | 0.4649 | 8.0655 | 4.3408 |
| 12.420 | 0.8000 | 1.3587 | 3.9924 | 3.7095 | 0.4654 | 8.0703 | 4.3397 |
| 24.233 | 0.7000 | 1.3294 | 4.0168 | 3.8361 | 0.4675 | 8.0761 | 4.3332 |
| 37.357 | 0.6000 | 1.2940 | 4.0360 | 3.9484 | 0.4687 | 8.0828 | 4.3240 |
| 52.620 | 0.5000 | 1.2813 | 4.0405 | 3.9792 | 0.4697 | 8.0905 | 4.3168 |
| 67.564 | 0.4000 | 1.2857 | 4.0363 | 3.9579 | 0.4701 | 8.0995 | 4.3150 |
| 82.033 | 0.3000 | 1.3030 | 4.0270 | 3.9033 | 0.4706 | 8.1104 | 4.3202 |
| 94.997 | 0.2000 | 1.3305 | 4.0157 | 3.8308 | 0.4708 | 8.1252 | 4.3372 |
| 107.306 | 0.1000 | 1.3741 | 4.0067 | 3.7511 | 0.4710 | 8.1481 | 4.3803 |
| 114.032 | 0.0500 | 1.4193 | 4.0042 | 3.6958 | 0.4712 | 8.1693 | 4.4321 |
| 117.879 | 0.0250 | 1.4781 | 3.9945 | 3.5982 | 0.4715 | 8.1880 | 4.4855 |
| 120.439 | 0.0100 | 1.5718 | 3.9631 | 3.3791 | 0.4721 | 8.2086 | 4.5565 |
| 121.939 | 0.0010 | 1.7950 | 3.7863 | 2.4484 | 0.4725 | 8.2436 | 4.7128 |
| 122.088 | 0.0001 | 1.8640 | 3.6975 | 2.0243 | 0.4725 | 8.2508 | 4.7724 |
| 122.143 | 0.0000 | 1.9002 | 3.6688 | 1.8735 | 0.4725 | 8.2509 | 4.8055 |
| 122.427 | 0.0000 | 2.0792 | 3.6100 | 1.4591 | 0.4729 | 8.2463 | 4.9753 |
| 123.352 | 0.0000 | 2.1040 | 3.6110 | 1.4385 | 0.4752 | 8.1974 | 5.1396 |
| 126.088 | 0.0000 | 1.9870 | 3.6931 | 1.8839 | 0.4779 | 8.1247 | 5.2331 |
| 137.096 | 0.0000 | 2.1216 | 3.8186 | 2.2514 | 0.4857 | 8.1041 | 5.4267 |
| 142.368 | 0.0000 | 2.3017 | 3.7395 | 1.7547 | 0.4897 | 8.0956 | 5.5894 |
| 144.667 | 0.0000 | 2.4688 | 3.6320 | 1.1577 | 0.4926 | 8.0837 | 5.7030 |

TABLE 2—*Continued*

| $t_6$ | $Y_c$ | $\log L/L_\odot$ | $\log T_{\rm eff}$ | $\log g_s$ | $M_c$ | $\log T_c$ | $\log \rho_c$ |
|---|---|---|---|---|---|---|---|
| 146.194 | 0.0000 | 2.6644 | 3.6016 | 0.8404 | 0.4967 | 8.0637 | 5.8053 |
| 146.956 | 0.0000 | 2.7937 | 3.7264 | 1.2104 | 0.5004 | 8.0493 | 5.8647 |
| 147.200 | 0.0000 | 2.8343 | 3.9217 | 1.9509 | 0.5020 | 8.0442 | 5.8845 |
| 147.377 | 0.0000 | 2.8599 | 4.1200 | 2.7183 | 0.5033 | 8.0401 | 5.8993 |
| 147.556 | 0.0000 | 2.8741 | 4.3194 | 3.5019 | 0.5047 | 8.0359 | 5.9147 |
| 147.743 | 0.0000 | 2.8578 | 4.5183 | 4.3139 | 0.5062 | 8.0313 | 5.9312 |
| 147.947 | 0.0000 | 2.7686 | 4.6944 | 5.1075 | 0.5075 | 8.0263 | 5.9506 |
| 148.174 | 0.0000 | 2.6046 | 4.8065 | 5.7196 | 0.5085 | 8.0207 | 5.9745 |
| 148.475 | 0.0000 | 2.4208 | 4.8844 | 6.2150 | 0.5091 | 8.0133 | 6.0100 |
| 148.584 | 0.0000 | 2.3916 | 4.9541 | 6.5233 | 0.5094 | 8.0139 | 6.0303 |
| 148.605 | 0.0000 | 2.1949 | 4.9475 | 6.6933 | 0.5094 | 8.0141 | 6.0343 |
| 148.638 | 0.0000 | 1.9956 | 4.9484 | 6.8965 | 0.5095 | 8.0143 | 6.0407 |
| 148.677 | 0.0000 | 1.7962 | 4.9343 | 7.0394 | 0.5095 | 8.0141 | 6.0479 |
| 148.744 | 0.0000 | 1.5972 | 4.9155 | 7.1630 | 0.5095 | 8.0129 | 6.0590 |
| 148.853 | 0.0000 | 1.3991 | 4.8882 | 7.2521 | 0.5095 | 8.0091 | 6.0749 |
| 149.033 | 0.0000 | 1.2016 | 4.8568 | 7.3241 | 0.5095 | 7.9994 | 6.0959 |
| 149.315 | 0.0000 | 1.0042 | 4.8248 | 7.3932 | 0.5095 | 7.9817 | 6.1220 |
| 149.695 | 0.0000 | 0.8070 | 4.7910 | 7.4552 | 0.5095 | 7.9580 | 6.1483 |
| 150.015 | 0.0000 | 0.6728 | 4.7669 | 7.4931 | 0.5095 | 7.9393 | 6.1652 |
| C. $M_{\rm env}^0 = 0.096\,M_\odot$ | | | | | | | |
| 1.220 | 0.9171 | 1.4990 | 3.6559 | 2.2637 | 0.4649 | 8.0645 | 4.3425 |
| 2.577 | 0.9000 | 1.5142 | 3.6523 | 2.2342 | 0.4649 | 8.0654 | 4.3428 |
| 12.162 | 0.8000 | 1.5339 | 3.6517 | 2.2119 | 0.4681 | 8.0706 | 4.3375 |
| 23.057 | 0.7000 | 1.5118 | 3.6625 | 2.2770 | 0.4715 | 8.0768 | 4.3272 |
| 35.164 | 0.6000 | 1.4907 | 3.6734 | 2.3416 | 0.4746 | 8.0840 | 4.3144 |
| 48.442 | 0.5000 | 1.4730 | 3.6846 | 2.4043 | 0.4777 | 8.0924 | 4.3035 |
| 61.450 | 0.4000 | 1.4612 | 3.6947 | 2.4566 | 0.4799 | 8.1017 | 4.2982 |
| 74.415 | 0.3000 | 1.4617 | 3.6995 | 2.4753 | 0.4816 | 8.1130 | 4.3011 |
| 86.056 | 0.2000 | 1.4728 | 3.6991 | 2.4626 | 0.4828 | 8.1280 | 4.3166 |
| 97.063 | 0.1000 | 1.5148 | 3.6877 | 2.3748 | 0.4839 | 8.1517 | 4.3600 |
| 103.028 | 0.0500 | 1.5695 | 3.6746 | 2.2677 | 0.4846 | 8.1730 | 4.4124 |
| 106.445 | 0.0250 | 1.6309 | 3.6608 | 2.1513 | 0.4853 | 8.1919 | 4.4664 |
| 108.649 | 0.0100 | 1.7054 | 3.6444 | 2.0113 | 0.4863 | 8.2121 | 4.5354 |
| 109.988 | 0.0010 | 1.8745 | 3.6150 | 1.7247 | 0.4869 | 8.2478 | 4.6892 |
| 110.122 | 0.0001 | 1.9349 | 3.6041 | 1.6203 | 0.4869 | 8.2561 | 4.7510 |
| 110.173 | 0.0000 | 1.9663 | 3.5986 | 1.5670 | 0.4868 | 8.2565 | 4.7866 |
| 110.480 | 0.0000 | 2.1643 | 3.5711 | 1.2590 | 0.4874 | 8.2540 | 4.9852 |
| 111.462 | 0.0000 | 2.1004 | 3.5817 | 1.3655 | 0.4899 | 8.1933 | 5.1402 |
| 122.339 | 0.0000 | 2.1475 | 3.5809 | 1.3153 | 0.4979 | 8.1147 | 5.4016 |
| 127.662 | 0.0000 | 2.3458 | 3.5561 | 1.0178 | 0.5014 | 8.1093 | 5.5980 |
| 129.838 | 0.0000 | 2.5441 | 3.5303 | 0.7165 | 0.5039 | 8.0936 | 5.7392 |
| 130.979 | 0.0000 | 2.7424 | 3.5047 | 0.4152 | 0.5072 | 8.0739 | 5.8397 |
| 131.862 | 0.0000 | 2.9406 | 3.4809 | 0.1219 | 0.5132 | 8.0486 | 5.9300 |
| 132.630 | 0.0000 | 3.1211 | 3.4627 | −0.1314 | 0.5226 | 8.0197 | 6.0122 |
| 133.287 | 0.0000 | 3.2780 | 3.4634 | −0.2854 | 0.5351 | 7.9936 | 6.0906 |
| 133.371 | 0.0000 | 3.2826 | 3.4665 | −0.2773 | 0.5371 | 7.9896 | 6.0998 |
| 133.414 | 0.0000 | 3.3256 | 3.4674 | −0.3170 | 0.5384 | 7.9806 | 6.0932 |
| 133.415 | 0.0000 | 3.1297 | 3.4869 | −0.0431 | 0.5384 | 7.9740 | 6.0872 |
| 133.415 | 0.0000 | 2.9339 | 3.5137 | 0.2601 | 0.5384 | 7.9738 | 6.0867 |
| 133.415 | 0.0000 | 2.7350 | 3.5348 | 0.5434 | 0.5382 | 7.9739 | 6.0868 |
| 133.415 | 0.0000 | 2.8083 | 3.5200 | 0.4108 | 0.5380 | 7.9741 | 6.0870 |
| 133.415 | 0.0000 | 3.0069 | 3.4968 | 0.1194 | 0.5379 | 7.9742 | 6.0870 |
| 133.415 | 0.0000 | 3.2046 | 3.4728 | −0.1744 | 0.5379 | 7.9743 | 6.0871 |
| 133.415 | 0.0000 | 3.4038 | 3.4592 | −0.4280 | 0.5379 | 7.9744 | 6.0872 |
| 133.415 | 0.0000 | 3.4512 | 3.4582 | −0.4793 | 0.5379 | 7.9744 | 6.0873 |

TABLE 3

Sample Metal-Poor Evolutionary Sequences:

[Fe/H] = −1.48, [O/Fe] = 0.60, $Y_{HB}$ = 0.25, $M_c$ = 0.485 $M_\odot$

| $t_6$ | $Y_c$ | $\log L/L_\odot$ | $\log T_{\rm eff}$ | $\log g_s$ | $M_c$ | $\log T_c$ | $\log \rho_c$ |
|---|---|---|---|---|---|---|---|
| \multicolumn{8}{c}{A. $M_{\rm env}^0 = 0.003\,M_\odot$} |
| 1.463 | 0.9500 | 1.1847 | 4.4470 | 5.6825 | 0.4855 | 8.0675 | 4.3201 |
| 5.540 | 0.9000 | 1.1949 | 4.4465 | 5.6705 | 0.4859 | 8.0697 | 4.3197 |
| 15.638 | 0.8000 | 1.2179 | 4.4453 | 5.6426 | 0.4861 | 8.0746 | 4.3147 |
| 28.403 | 0.7000 | 1.2420 | 4.4437 | 5.6120 | 0.4852 | 8.0804 | 4.3033 |
| 42.790 | 0.6000 | 1.2686 | 4.4424 | 5.5801 | 0.4852 | 8.0871 | 4.2927 |
| 56.334 | 0.5000 | 1.2962 | 4.4416 | 5.5494 | 0.4852 | 8.0947 | 4.2859 |
| 68.835 | 0.4000 | 1.3249 | 4.4415 | 5.5203 | 0.4852 | 8.1037 | 4.2837 |
| 79.881 | 0.3000 | 1.3554 | 4.4425 | 5.4938 | 0.4852 | 8.1147 | 4.2885 |
| 90.512 | 0.2000 | 1.3875 | 4.4455 | 5.4741 | 0.4852 | 8.1293 | 4.3050 |
| 101.598 | 0.1000 | 1.4213 | 4.4539 | 5.4736 | 0.4852 | 8.1528 | 4.3490 |
| 107.804 | 0.0500 | 1.4379 | 4.4641 | 5.4977 | 0.4852 | 8.1739 | 4.4015 |
| 111.370 | 0.0250 | 1.4451 | 4.4746 | 5.5326 | 0.4852 | 8.1924 | 4.4551 |
| 113.709 | 0.0100 | 1.4509 | 4.4877 | 5.5792 | 0.4861 | 8.2128 | 4.5240 |
| 115.024 | 0.0010 | 1.4708 | 4.5162 | 5.6731 | 0.4864 | 8.2501 | 4.6797 |
| 115.157 | 0.0001 | 1.4903 | 4.5265 | 5.6951 | 0.4864 | 8.2592 | 4.7446 |
| 115.199 | 0.0000 | 1.4996 | 4.5309 | 5.7031 | 0.4864 | 8.2601 | 4.7755 |
| 116.426 | 0.0000 | 1.5928 | 4.5714 | 5.7720 | 0.4872 | 8.2064 | 5.1502 |
| 119.450 | 0.0000 | 1.7927 | 4.5764 | 5.5922 | 0.4852 | 8.1434 | 5.2943 |
| 126.761 | 0.0000 | 1.9914 | 4.5950 | 5.4677 | 0.4851 | 8.1019 | 5.4648 |
| 132.896 | 0.0000 | 2.1849 | 4.6420 | 5.4626 | 0.4851 | 8.0744 | 5.6876 |
| 136.420 | 0.0000 | 2.3122 | 4.7569 | 5.7949 | 0.4855 | 8.0170 | 5.8972 |
| 136.702 | 0.0000 | 2.4994 | 4.7949 | 5.7596 | 0.4861 | 8.0138 | 5.9230 |
| 136.901 | 0.0000 | 2.4189 | 4.8624 | 6.1100 | 0.4868 | 8.0139 | 5.9481 |
| 136.958 | 0.0000 | 2.2191 | 4.8637 | 6.3149 | 0.4870 | 8.0125 | 5.9534 |
| 136.984 | 0.0000 | 2.0200 | 4.8458 | 6.4425 | 0.4870 | 8.0108 | 5.9539 |
| 136.991 | 0.0000 | 1.8254 | 4.8024 | 6.4632 | 0.4870 | 8.0014 | 5.9388 |
| 136.991 | 0.0000 | 1.7369 | 4.7705 | 6.4242 | 0.4870 | 7.9958 | 5.9294 |
| \multicolumn{8}{c}{B. $M_{\rm env}^0 = 0.035\,M_\odot$} |
| 1.252 | 0.9500 | 1.2535 | 4.2880 | 5.0054 | 0.4852 | 8.0700 | 4.2971 |
| 4.922 | 0.9000 | 1.2614 | 4.2856 | 4.9879 | 0.4852 | 8.0722 | 4.2970 |
| 13.960 | 0.8000 | 1.2814 | 4.2801 | 4.9457 | 0.4852 | 8.0771 | 4.2928 |
| 25.506 | 0.7000 | 1.3024 | 4.2737 | 4.8993 | 0.4852 | 8.0828 | 4.2821 |
| 38.633 | 0.6000 | 1.3264 | 4.2669 | 4.8481 | 0.4852 | 8.0896 | 4.2722 |
| 51.086 | 0.5000 | 1.3516 | 4.2602 | 4.7958 | 0.4852 | 8.0972 | 4.2662 |
| 62.542 | 0.4000 | 1.3783 | 4.2536 | 4.7431 | 0.4852 | 8.1061 | 4.2648 |
| 72.848 | 0.3000 | 1.4071 | 4.2475 | 4.6899 | 0.4852 | 8.1172 | 4.2703 |
| 82.024 | 0.2000 | 1.4374 | 4.2431 | 4.6418 | 0.4852 | 8.1320 | 4.2878 |
| 91.391 | 0.1000 | 1.4702 | 4.2439 | 4.6124 | 0.4852 | 8.1552 | 4.3330 |
| 96.655 | 0.0500 | 1.4881 | 4.2512 | 4.6237 | 0.4852 | 8.1763 | 4.3855 |
| 99.801 | 0.0250 | 1.4982 | 4.2614 | 4.6545 | 0.4852 | 8.1950 | 4.4390 |
| 101.790 | 0.0100 | 1.5106 | 4.2744 | 4.6936 | 0.4852 | 8.2154 | 4.5077 |
| 103.006 | 0.0010 | 1.5695 | 4.2939 | 4.7131 | 0.4852 | 8.2523 | 4.6610 |
| 103.128 | 0.0001 | 1.6124 | 4.2934 | 4.6679 | 0.4852 | 8.2611 | 4.7230 |
| 103.174 | 0.0000 | 1.6389 | 4.2911 | 4.6321 | 0.4852 | 8.2623 | 4.7587 |
| 103.395 | 0.0000 | 1.8336 | 4.2472 | 4.2622 | 0.4852 | 8.2597 | 4.9297 |
| 103.651 | 0.0000 | 2.0139 | 4.1617 | 3.7394 | 0.4852 | 8.2572 | 5.0469 |
| 112.399 | 0.0000 | 2.0914 | 4.1419 | 3.5831 | 0.4890 | 8.1127 | 5.4209 |
| 116.764 | 0.0000 | 2.2518 | 4.0235 | 2.9490 | 0.4908 | 8.1029 | 5.5662 |
| 118.893 | 0.0000 | 2.3793 | 3.8698 | 2.2069 | 0.4923 | 8.0909 | 5.6695 |
| 120.191 | 0.0000 | 2.4946 | 3.7082 | 1.4450 | 0.4940 | 8.0768 | 5.7498 |
| 121.671 | 0.0000 | 2.6886 | 3.6659 | 1.0818 | 0.4980 | 8.0490 | 5.8586 |
| 122.923 | 0.0000 | 2.8880 | 3.6591 | 0.8551 | 0.5053 | 8.0124 | 5.9571 |
| 123.271 | 0.0000 | 3.0068 | 3.6536 | 0.7143 | 0.5083 | 8.0036 | 5.9889 |
| 123.319 | 0.0000 | 2.9525 | 3.7979 | 1.3461 | 0.5119 | 7.9807 | 5.9570 |
| 123.319 | 0.0000 | 2.8852 | 3.9860 | 2.1656 | 0.5119 | 7.9797 | 5.9552 |
| 123.319 | 0.0000 | 2.7668 | 4.1447 | 2.9188 | 0.5119 | 7.9792 | 5.9544 |
| 123.320 | 0.0000 | 2.5817 | 4.2108 | 3.3685 | 0.5118 | 7.9791 | 5.9543 |
| 123.320 | 0.0000 | 2.3979 | 4.1472 | 3.2979 | 0.5088 | 7.9793 | 5.9545 |
| 123.320 | 0.0000 | 2.4100 | 3.9592 | 2.5337 | 0.5085 | 7.9795 | 5.9549 |
| 123.320 | 0.0000 | 2.4889 | 3.7757 | 1.7207 | 0.5085 | 7.9797 | 5.9551 |
| 123.320 | 0.0000 | 2.6201 | 3.6572 | 1.1153 | 0.5083 | 7.9798 | 5.9553 |
| 123.320 | 0.0000 | 2.8193 | 3.6399 | 0.8471 | 0.5082 | 7.9799 | 5.9554 |
| 123.320 | 0.0000 | 3.0185 | 3.6227 | 0.5791 | 0.5082 | 7.9800 | 5.9556 |
| 123.320 | 0.0000 | 3.2181 | 3.6108 | 0.3317 | 0.5081 | 7.9801 | 5.9559 |
| 123.322 | 0.0000 | 3.2812 | 3.6084 | 0.2594 | 0.5080 | 7.9837 | 5.9617 |
| 123.324 | 0.0000 | 3.0823 | 3.6243 | 0.5217 | 0.5082 | 7.9861 | 5.9656 |
| 123.332 | 0.0000 | 2.9165 | 3.6669 | 0.8581 | 0.5085 | 7.9837 | 5.9616 |
| 123.350 | 0.0000 | 2.9065 | 3.7392 | 1.1574 | 0.5088 | 7.9867 | 5.9675 |

TABLE 3—*Continued*

| $t_6$ | $Y_c$ | $\log L/L_\odot$ | $\log T_{\rm eff}$ | $\log g_s$ | $M_c$ | $\log T_c$ | $\log \rho_c$ |
|---|---|---|---|---|---|---|---|
| 123.398 | 0.0000 | 3.0572 | 3.6354 | 0.5912 | 0.5094 | 7.9952 | 5.9855 |
| 123.599 | 0.0000 | 3.2474 | 3.6372 | 0.4083 | 0.5136 | 8.0056 | 6.0324 |
| 123.624 | 0.0000 | 3.1687 | 3.8079 | 1.1696 | 0.5142 | 7.9970 | 6.0219 |
| 123.625 | 0.0000 | 3.1419 | 4.0061 | 1.9893 | 0.5142 | 7.9927 | 6.0147 |
| 123.625 | 0.0000 | 3.0957 | 4.2004 | 2.8127 | 0.5142 | 7.9889 | 6.0082 |
| 123.625 | 0.0000 | 3.0046 | 4.3774 | 3.6119 | 0.5142 | 7.9848 | 6.0015 |
| 123.625 | 0.0000 | 2.8532 | 4.5058 | 4.2768 | 0.5142 | 7.9814 | 5.9956 |
| 123.626 | 0.0000 | 2.6630 | 4.5616 | 4.6905 | 0.5142 | 7.9793 | 5.9916 |
| 123.627 | 0.0000 | 2.7527 | 4.4467 | 4.1408 | 0.5142 | 7.9794 | 5.9912 |
| 123.628 | 0.0000 | 2.9014 | 4.3135 | 3.4594 | 0.5141 | 7.9800 | 5.9919 |
| 123.629 | 0.0000 | 2.9955 | 4.2174 | 2.9810 | 0.5141 | 7.9825 | 5.9954 |
| 123.635 | 0.0000 | 2.9209 | 4.3895 | 3.7440 | 0.5142 | 7.9874 | 6.0029 |
| 123.645 | 0.0000 | 3.0451 | 4.2347 | 3.0005 | 0.5142 | 7.9923 | 6.0112 |
| 123.656 | 0.0000 | 3.1229 | 4.0507 | 2.1871 | 0.5144 | 7.9953 | 6.0173 |
| 123.669 | 0.0000 | 3.1796 | 3.8590 | 1.3631 | 0.5146 | 7.9981 | 6.0238 |
| 123.723 | 0.0000 | 3.2888 | 3.7632 | 0.8709 | 0.5159 | 8.0041 | 6.0434 |
| 123.750 | 0.0000 | 3.3160 | 3.9608 | 1.6340 | 0.5166 | 8.0056 | 6.0512 |
| 123.768 | 0.0000 | 3.3283 | 4.1604 | 2.4200 | 0.5172 | 8.0063 | 6.0558 |
| 123.788 | 0.0000 | 3.3359 | 4.3602 | 3.2118 | 0.5177 | 8.0068 | 6.0606 |
| 123.809 | 0.0000 | 3.3298 | 4.5601 | 4.0171 | 0.5184 | 8.0072 | 6.0656 |
| 123.830 | 0.0000 | 3.2933 | 4.7561 | 4.8378 | 0.5190 | 8.0075 | 6.0704 |
| 123.850 | 0.0000 | 3.1798 | 4.9172 | 5.5956 | 0.5195 | 8.0076 | 6.0746 |
| 123.863 | 0.0000 | 2.9970 | 4.9948 | 6.0889 | 0.5197 | 8.0076 | 6.0774 |
| 123.871 | 0.0000 | 2.7998 | 5.0262 | 6.4118 | 0.5198 | 8.0075 | 6.0790 |
| 123.874 | 0.0000 | 2.6000 | 5.0216 | 6.5930 | 0.5198 | 8.0075 | 6.0796 |
| 123.874 | 0.0000 | 2.4010 | 5.0018 | 6.7129 | 0.5198 | 8.0076 | 6.0798 |
| 123.875 | 0.0000 | 2.2027 | 4.9759 | 6.8078 | 0.5198 | 8.0076 | 6.0800 |
| 123.876 | 0.0000 | 2.0049 | 4.9463 | 6.8873 | 0.5198 | 8.0077 | 6.0803 |
| 123.878 | 0.0000 | 1.8074 | 4.9146 | 6.9580 | 0.5198 | 8.0077 | 6.0808 |
| 123.919 | 0.0000 | 1.6095 | 4.8865 | 7.0433 | 0.5198 | 8.0064 | 6.0874 |
| 124.185 | 0.0000 | 1.4106 | 4.8780 | 7.2082 | 0.5199 | 7.9921 | 6.1213 |
| 124.391 | 0.0000 | 1.2121 | 4.8535 | 7.3087 | 0.5199 | 7.9789 | 6.1413 |
| 124.653 | 0.0000 | 1.0141 | 4.8250 | 7.3928 | 0.5199 | 7.9623 | 6.1617 |
| 124.988 | 0.0000 | 0.8166 | 4.7936 | 7.4646 | 0.5199 | 7.9422 | 6.1819 |
| 125.419 | 0.0000 | 0.6196 | 4.7592 | 7.5238 | 0.5199 | 7.9189 | 6.2010 |
| 125.968 | 0.0000 | 0.4231 | 4.7223 | 7.5732 | 0.5199 | 7.8933 | 6.2183 |
| 126.666 | 0.0000 | 0.2269 | 4.6835 | 7.6140 | 0.5199 | 7.8668 | 6.2331 |
| 127.539 | 0.0000 | 0.0310 | 4.6430 | 7.6478 | 0.5199 | 7.8406 | 6.2455 |
| 127.785 | 0.0000 | −0.0164 | 4.6329 | 7.6550 | 0.5199 | 7.8343 | 6.2482 |
| C. $M_{\rm env}^0 = 0.105\,M_\odot$ | | | | | | | |
| 1.231 | 0.9500 | 1.5664 | 4.0152 | 3.6561 | 0.4852 | 8.0710 | 4.2887 |
| 4.701 | 0.9000 | 1.5801 | 4.0068 | 3.6089 | 0.4852 | 8.0732 | 4.2880 |
| 13.253 | 0.8000 | 1.5852 | 4.0100 | 3.6164 | 0.4861 | 8.0785 | 4.2803 |
| 23.895 | 0.7000 | 1.5739 | 4.0266 | 3.6941 | 0.4892 | 8.0850 | 4.2644 |
| 35.617 | 0.6000 | 1.5580 | 4.0440 | 3.7801 | 0.4913 | 8.0923 | 4.2496 |
| 46.772 | 0.5000 | 1.5499 | 4.0528 | 3.8232 | 0.4932 | 8.1001 | 4.2401 |
| 57.383 | 0.4000 | 1.5519 | 4.0533 | 3.8229 | 0.4942 | 8.1098 | 4.2367 |
| 66.439 | 0.3000 | 1.5636 | 4.0473 | 3.7874 | 0.4954 | 8.1208 | 4.2413 |
| 74.770 | 0.2000 | 1.5854 | 4.0367 | 3.7232 | 0.4958 | 8.1361 | 4.2590 |
| 82.958 | 0.1000 | 1.6263 | 4.0194 | 3.6129 | 0.4960 | 8.1594 | 4.3054 |
| 87.699 | 0.0500 | 1.6727 | 3.9988 | 3.4843 | 0.4966 | 8.1805 | 4.3600 |
| 90.575 | 0.0250 | 1.7268 | 3.9666 | 3.3014 | 0.4969 | 8.1991 | 4.4162 |
| 92.531 | 0.0100 | 1.7956 | 3.9103 | 3.0074 | 0.4975 | 8.2191 | 4.4859 |
| 93.606 | 0.0010 | 1.9425 | 3.7098 | 2.0584 | 0.4981 | 8.2551 | 4.6438 |
| 93.712 | 0.0001 | 1.9877 | 3.6907 | 1.9370 | 0.4981 | 8.2636 | 4.7047 |
| 93.751 | 0.0000 | 2.0107 | 3.6847 | 1.8897 | 0.4977 | 8.2646 | 4.7380 |
| 94.062 | 0.0000 | 2.2090 | 3.6596 | 1.5912 | 0.4982 | 8.2656 | 4.9730 |
| 99.393 | 0.0000 | 2.1465 | 3.6697 | 1.6941 | 0.5052 | 8.1348 | 5.3386 |
| 105.474 | 0.0000 | 2.3455 | 3.6501 | 1.4167 | 0.5081 | 8.1223 | 5.5746 |
| 107.951 | 0.0000 | 2.5448 | 3.6342 | 1.1539 | 0.5100 | 8.1033 | 5.7526 |
| 109.171 | 0.0000 | 2.7442 | 3.6203 | 0.8986 | 0.5130 | 8.0759 | 5.8766 |
| 110.059 | 0.0000 | 2.9437 | 3.6063 | 0.6435 | 0.5186 | 8.0408 | 5.9751 |
| 110.505 | 0.0000 | 2.9985 | 3.6108 | 0.6065 | 0.5230 | 8.0173 | 6.0197 |
| 110.607 | 0.0000 | 3.1341 | 3.5991 | 0.4239 | 0.5247 | 8.0172 | 6.0399 |
| 110.608 | 0.0000 | 2.9501 | 3.6155 | 0.6736 | 0.5247 | 7.9970 | 6.0055 |
| 110.608 | 0.0000 | 2.7507 | 3.6313 | 0.9362 | 0.5246 | 7.9972 | 6.0058 |
| 110.609 | 0.0000 | 2.6880 | 3.6332 | 1.0067 | 0.5244 | 7.9975 | 6.0062 |
| 110.609 | 0.0000 | 2.8874 | 3.6177 | 0.7451 | 0.5243 | 7.9976 | 6.0064 |
| 110.609 | 0.0000 | 3.0867 | 3.6017 | 0.4815 | 0.5243 | 7.9977 | 6.0066 |
| 110.609 | 0.0000 | 3.2805 | 3.5917 | 0.2477 | 0.5243 | 7.9978 | 6.0068 |
| 110.609 | 0.0000 | 3.2823 | 3.5852 | 0.2202 | 0.5242 | 7.9978 | 6.0067 |
| 110.610 | 0.0000 | 3.4167 | 3.5720 | 0.0329 | 0.5240 | 8.0001 | 6.0106 |

TABLE 3—*Continued*

| $t_6$ | $Y_c$ | $\log L/L_\odot$ | $\log T_{\text{eff}}$ | $\log g_s$ | $M_c$ | $\log T_c$ | $\log \rho_c$ |
|---|---|---|---|---|---|---|---|
| 110.611 | 0.0000 | 3.2187 | 3.5916 | 0.3094 | 0.5243 | 8.0022 | 6.0139 |
| 110.618 | 0.0000 | 3.0196 | 3.6094 | 0.5796 | 0.5246 | 8.0003 | 6.0111 |
| 110.659 | 0.0000 | 3.1601 | 3.5974 | 0.3911 | 0.5251 | 8.0095 | 6.0314 |
| 110.787 | 0.0000 | 3.3586 | 3.5772 | 0.1120 | 0.5284 | 8.0206 | 6.0751 |
| 110.887 | 0.0000 | 3.2144 | 3.5952 | 0.3280 | 0.5318 | 8.0014 | 6.0667 |
| 110.889 | 0.0000 | 3.0153 | 3.6125 | 0.5965 | 0.5318 | 7.9928 | 6.0514 |
| 110.891 | 0.0000 | 3.0182 | 3.6116 | 0.5901 | 0.5317 | 7.9935 | 6.0520 |
| 110.906 | 0.0000 | 3.2041 | 3.5957 | 0.3402 | 0.5318 | 8.0036 | 6.0694 |
| 110.969 | 0.0000 | 3.4027 | 3.5758 | 0.0623 | 0.5337 | 8.0144 | 6.0999 |
| 111.018 | 0.0000 | 3.4503 | 3.5744 | 0.0091 | 0.5355 | 8.0162 | 6.1152 |

## *References*

Alexander, D.R. 1975, ApJS, 29, 363

Brocato, E., Matteuci, F., Mazzitelli, I., and Tornambè 1990, ApJ, 349, 458

Burstein, D., Bertola, F., Buson, L.M., Faber, S.M. and Lauer, T.R. 1988, ApJ, 328, 440

Caloi, V. 1989, A&A, 221, 27

Caloi, V. 1987, Castellani, V., and Piccolo, F., A&AS, 67, 181

Castellani, M. and Tornambè, A. 1991, ApJ, 381, 393

Castellani, M., Limongi, M. and Tornambè, A. 1992, ApJ, 389, 227

Castellani, V., Chieffi, A., Pulone, L., and Tornambè, A. 1985, ApJ, 296, 204

Caughlan, G.R., and Fowler, W.A. 1988, *Atomic and Nuclear Data Tables,* 40, 283

Ciardullo, R.B. and Demarque, P. 1978 in *The HR Diagram* eds. A.G.D. Philip and D.S. Hayes (Reidel: Dordrecht), p. 345

Clegg, R.E.S. and Middlemass 1987, MNRAS, 228, 759

Code, A.D. 1969, PASP, 81, 475

de Boer, K.S. 1985, A&A, 142, 321

Dorman, B. 1990 *Ph.D. thesis, University of Victoria*

Dorman, B. 1992a, ApJS, 81, 221

Dorman, B. 1992b, ApJS, 80, 701

Dorman, B. and Rood, R.T, 1993, ApJ, 409, 387

Dupree, A. K. 1986, ARAA, 24, 377.

Eggleton, P.P., Faulkner, J., and Flannery, B.F. 1973, A&A, 23, 325

Faber, S.M. 1983, *Highlights of Astronomy,* 6, 165

Ferguson, H.C. & Davidsen, A.F. 1993, ApJ , *in press.*

Fujimoto, M.Y. and Iben, I. Jr 1991, ApJ, 374, 631

Fusi Pecci, F., Ferraro, F.R., Corsi, C.E., Cacciari, C., and Buonanno, R. 1992, AJ, 104, 1831

Gingold, R.A. 1976, ApJ, 204, 116

Green, R.F., Schmidt, M., and Liebert, J.W. 1986, ApJS, 61, 305

Greenstein, J.L., and Sargent, A.I. 1974, ApJS, 28, 157

Greggio, L. and Renzini, A. 1990, ApJ, 364, 35

Heber, U. 1987 in *The Second Conference on Faint Blue Stars* eds. A.G.D. Philip, D.S. Hayes & J.W. Leibert (Schenectady: Davis), p. 79

Horch, E., Demarque, P. and Pinnsoneault, M. 1992, ApJ. L. , 388, L53 (HDP)

Huebner, W.F., Merts, A.L., Magee, N.H., and Argo M.F. 1977 *Los Alamos Sci. Lab. Rep.* No. LA-6760-M

Iben, I., Kaler, J., Truran, J. and Renzini, A. 1983, ApJ, 264, 605

Iben, I. and Rood, R.T. 1970, ApJ, 161, 587

Krishna-Swamy, K.S. 1966, ApJ, 145, 174

Kurucz, R.L. 1991 in in *Precision Photometry: Astrophysics of the Galaxy* eds. A.G.D. Philip, A.R. Upgren, & K.A. Janes (Schenectady: Davis), p. 27

Landsman, W.B. *et al.* 1992, ApJ. L. , 395, L21

Lee, Y-W., Demarque, P., and Zinn, R.J. 1990, ApJ, 350, 155

MacGregor, K.B. and Stencel, R.E. 1992, ApJ, 397, 644

O'Connell, R.W. 1993 in *The Globular Cluster-Galaxy Connection* eds. J.P. Brodie and G.H. Smith (ASP: San Francisco), in press.

Pacyzński 1971, Acta Astr, 21, 417

Pagel, B. E. J., Simonson, E. A., Terlevich, R. J. & Edmunds, M. G. 1992, MNRAS, 255, 325

Peimbert, M., Torres-Peimbert, S., & Ruiz, M. T. 1992, Rev. Mexicana A&A, 24, 155

Reimers, D. 1975, Mém. Soc. Roy. Sci. Liège, $6^e$ Ser., 8, 369

Renzini, A. 1981a in *Physical Processes in Red Giants* eds I. Iben Jr and A. Renzini (Dordrecht: Reidel), p. 431

Renzini, A. 1981b in *Effects of Mass Loss on Stellar Evolution* eds C. Chiosi and R. Stalio (Dordrecht: Reidel), p. 319

Renzini A. and Fusi Pecci, F. 1988, ARAA, 26, 199.

Renzini, A. and Buzzoni, A. 1986 in *Spectral Evolution of Galaxies* eds C. Chiosi and A. Renzini (Dordrecht: Reidel), p. 195

Rood, R.T. 1973, ApJ, 184, 815

Schönberner, D. 1979, A&A, 79, 108

Schönberner, D. 1983, ApJ, 272, 708

Sweigart, A.V. and Gross, P.G. 1976, ApJS, 32, 367

Skillman, E. D. 1993 in *Proceedings of the 16th Texas Symposium on Relativistic Astrophysics and 3rd Symposium on Particles, Strings, and Cosmology* eds. C. Akerlof, M. Srednicki (NYAS: New York), in press.

Tinsley 1980, Fund. Cos. Phys., 5, 287

Tripicco, M.J., Dorman, B., and Bell, R.A. 1993, AJ, *in press*

van Albada, T.S., de Boer, K.S., and Dickens, R.J. 1981, MNRAS, 195, 591

VandenBerg, D.A. 1992, ApJ, 391, 685

Welch, G.A. and Code, A.D. 1979, ApJ, 236, 798

Wood, P.R. and Faulkner, D.J. 1986, ApJ, 307, 659

Wood, P.R. and Zarro, D.M. 1981, ApJ, 247, 247

Zinn, R.J., Newell, E.B., and Gibson, J.B. 1972, A&A, 18, 390

## *Figure Captions*

Fig. 1  Schematic representation of the different types of post-HB evolutionary sequences described in the text. "EHB" stars have envelope masses that are too small to reach the end stages of normal AGB evolution. Also marked are the places in the HR diagram where other physical processes are important.

Fig. 2  Zero Age HB sequences for models with varying abundance. (a) Models with [Fe/H] $\leq 0$, and with $Y$ determined from a linear relation for $\Delta Y/\Delta Z$ such that $Y_{ZAMS} = 0.235, 0.27$ at [Fe/H] $= -2.3, 0$ respectively. Helium added to the envelope during the first dredge-up phase is included. The least massive model has $M_{env} = 0.002 - 0.003\,M_\odot$ in each case. (b) ZAHBs with [Fe/H] $= 0.39$ ($Z = 0.04$) and [Fe/H] $= 0.58$ ($Z = 0.06$) and with two choices each for $Y$. The lower value corresponds to no additional helium above the solar value for [Fe/H] $> 0$. The other choice is $\Delta Y/\Delta Z = 3$ for [Fe/H] $= 0.39$ and $\Delta Y/\Delta Z = 4$ for [Fe/H] $= 0.58$.

Fig. 3  Evolutionary tracks for eight different compositions corresponding to the ZAHB sequences illustrated in Figure 2. In each of the panels, filled circles mark the horizontal-branch phase of evolution, and they are placed at intervals of 10 Myr. The crosses represent the core helium exhaustion point ($Y_c \leq 0.0001$). Open circles mark the post-HB stages, and are placed at intervals of 2 Myr. For the part of the sequence after the luminosity has reached 1000 $L_\odot$, filled triangles mark intervals of 100,000 Myr. Other input parameters are listed in Table 1. (a) [Fe/H] $= -2.26$, [O/Fe] $= 0.50$, $M = 0.498$ to $0.90\,M_\odot$ (b) [Fe/H] $= -1.48$, [O/Fe] $= 0.63$, $M = 0.488$ to $0.90\,M_\odot$ (c) [Fe/H] $= -0.47$, [O/Fe] $= 0.23$, $M = 0.478$ to $0.90\,M_\odot$ (d) [Fe/H] $= 0$, $M = 0.471$ to $0.90\,M_\odot$, $Y = 0.29$ (e) [Fe/H] $= 0.39$, $M = 0.467$ to $0.80\,M_\odot$, $Y = 0.29$ (f) [Fe/H] $= 0.58$, $M = 0.461$ to $0.90\,M_\odot$, $Y = 0.36$ (g) [Fe/H] $= 0.39$, $M = 0.457$ to $0.90\,M_\odot$, $Y = 0.29$ (h) [Fe/H] $= 0.58$, $M = 0.440$ to $0.80\,M_\odot$, $Y = 0.46$.

Fig. 4  Tracks from Fig. 3 illustrating each of the classes of post-HB morphology discussed in the text. Dashed-dotted line: AGB-sequence, evolving to higher luminosity, which will eventually cross the HR diagram as a P-AGB star. Dotted line: Post-Early AGB sequence. Solid line: AGB-Manqué sequence. (a) Metal rich sequences from Fig. 3e. (b) Metal-Poor sequences from Fig. 3b. The AGB-Manqué sequence is terminated during a thermal pulse. The P-EAGB model undergoes a thermal pulse on its way to the cooling sequence at the left of the diagram.

Fig. 5  Evolutionary tracks illustrating the rapid change in luminosity and evolutionary timescale near the transition mass, $M_{env}^{TP}$. The sequences have [Fe/H] $= 0$ and $M = 0.50, 0.51, 0.52,$ and $0.53$ $M_\odot$. Filled triangles are placed at intervals of 10,000 yr in the upper two sequences. For the lower sequences, open circles mark intervals of 250,000 years of evolution.

Fig. 6  Integrated energy in a band centered close to 1500 Å for tracks of solar metallicity, plotted as a function of $M_{env}$. The energy is given in units of $10^{48}$ erg Å$^{-1}$ per star. The points corresponding to sequences from Figure 5 are labelled.

Fig. 7  Estimates of the mass lost from evolving sequences, derived from the Reimers formula applied to constant mass evolutionary tracks. The ordinate, given on a logarithmic scale, is in units $1/\eta\,M_\odot$, where $\eta$ is the mass loss parameter. The masses plotted are 0.51 $M_\odot$ (post-Early AGB model), and 0.53, 0.60, 0.80 $M_\odot$.